\documentclass{aa}
\usepackage[varg]{txfonts}

\usepackage{graphicx}
\usepackage{natbib}
\usepackage{xspace}

\def\gray{$\gamma$-ray\xspace}
\def\grays{$\gamma$-rays\xspace}
\def\fermi{{\it Fermi}-LAT\xspace}

\def\deg{\hbox{$^\circ$}}

\begin{document}

\title{The diffuse gamma-ray emission  toward the Galactic mini starburst W43 }
\titlerunning{gamma-rays toward W43}
\author{Rui-Zhi Yang\inst{1,2,3}
\and Yuan Wang\inst{4}
}
\institute{Department of Astronomy, School of Physical Sciences, University of Science and Technology of China, Hefei, Anhui 230026, China
\and  CAS Key Labrotory for Research in Galaxies and Cosmology, University of Science and Technology of China, Hefei, Anhui 230026, China 
\and School of Astronomy and Space Science, University of Science and Technology of China, Hefei, Anhui 230026, China 
\and Max-Planck-Institut f\"ur Astronomie,  Königstuhl 17, 69117 Heidelberg, Germany
}

\abstract {
In this paper we report the Fermi Large Area Telescope (\fermi) detection of the \gray emission toward the young star forming region W43.
Using the latest source catalog and diffuse background models, the extended \gray excess is detected with a significance of $\sim 16 \sigma$. The \gray emission has a spectrum with a photon index of $2.3 \pm 0.1$. 
We also performed a detailed analysis of the gas content in this region by taking into account the opacity correction to the HI gas column density. 
The total cosmic-ray (CR) proton energy is estimated to be on the order of $10^{48}\ \rm erg,$ assuming the \grays are produced from the interaction of the accelerated protons and nuclei with the ambient gas. Comparing this region to the other star formation regions in our Galaxy, we find that the CR luminosity is better correlated with the wind power than the star formation rate (SFR). This result suggests that CRs are primarily accelerated by stellar wind in these systems.}
 \maketitle

\section{Introduction}
The star formation processes are strongly related to cosmic rays (CRs).  On the one hand, the  CRs are believed to be accelerated by shocks in supernova remnants \citep[SNRs; see, e.g.,][]{Blasi13} or massive star wind \citep{aharonian19}, thus star formation is the ultimate energy source of CRs. On the other hand, CRs control the ionization rate in the core of dense molecular clouds and play an important role in determining the Jeans mass and initial mass function (IMF) of stars \citep{papadopoulos10}. On the galactic scale, the derived CR densities are strongly correlated with the star formation rate (SFR) \citep{fermi_galaxy}. 
In this regard, W43 is an ideal location to study the relation between star formation and CRs. W43 is known as a Galactic mini-starburst region \citep{motte03} and contributes about 5-10\% of the SFR in the entire Milky Way \citep{luong11}. The center of W43 is the large \ion{H}{ii} region  W43-main, which is excited by a Wolf-Rayet and OB star cluster. Some \gray emissions have also been detected in this region and an extended TeV source has been detected by the HESS telescope \cite{hgps}. \citet{fermi_pwn} have also found GeV excess in this region and attribute it to a potential pulsar wind nebula (PWN), but the powering pulsar has not yet been found. With the improved understanding of the Fermi Large Area Telescope (\fermi) instrument response and the accumulation of data, it is interesting to revisit the Fermi-LAT observations in this region. Furthermore, \citet{bihr15} performed a state-of-the-art analysis of the gas content of this region, and find the optical depth correction is significant for the HI gas column density.  Thus it is possible to derive the CR distribution with an unprecedented accuracy.

%

In this paper, we perform a detailed analysis on the ten-year Fermi-LAT data on this region and investigate the gas and CR content therein. The paper is organized as follows. In Sect. 2 we present the details of the data analysis. In Sect. 3 we investigate the gas distribution in this region. In Sect. 4 we discuss the  possible radiation mechanisms of the \gray emission in this region. In Sect. 5 we discuss the implications of our observations. 
%
%


%
%



\section{\fermi data analysis}
We selected the \fermi Pass 8 database on the W43 region from August 4, 2008 (MET 239557417), to April 3, 2019 (MET 575964102), for the analysis.
 To take advantage of a better angular resolution, we discarded PSF 0 event types in the analysis.   We chose an A 10\deg $\times$ 10\deg square region centered at the position of W43 (R.A. = 281.885\deg, Dec. = -1.942\deg)  as the region of interest (ROI).
We used the"'source" event class and the recommended data cut expression $\rm (DATA\_QUAL > 0) \&\& (LAT\_CONFIG == 1)$ to exclude time periods when spacecraft events affected the data quality.
To reduce the background contamination from the Earth's albedo, only the events with zenith angles less than 90\deg are included in the analysis.
We used the  Fermitools from Conda distribution\footnote{https://github.com/fermi-lat/Fermitools-conda/} together with the latest version of the instrument response functions (IRFs) {\it P8R3\_SOURCE\_V2} in the analysis.

In our background model, we included the sources in the \fermi eight-year catalog \citep[4FGL,][]{Fermi19} within the ROI, enlarged by 7\deg.
We left the normalizations and spectral indices free for all sources within 8\deg\  of W43.
For the diffuse background components, we used the latest Galactic diffuse model {\it gll\_iem\_v07.fits} and isotropic emission model {\it iso\_P8R3\_SOURCE\_V2\_v1.txt}\footnote{\url{https://fermi.gsfc.nasa.gov/ssc/data/access/lat/BackgroundModels.html}} with their normalization parameters free.

\begin{table*}[htbp]
\caption{Fitting results for different models.} \label{tab:loglike} \centering
\begin{tabular}{llll}
\hline
Model &\vline ~-log(likelihood)&\vline ~free parameters&\vline ~AIC \\
\hline
 4FGL catalog - unassociated point sources   &\vline ~76491 &\vline ~103&\vline ~153188\\

\hline
 4FGL catalog   &\vline  ~76379 &\vline ~112&\vline ~152982\\
\hline
 4FGL catalog - unassociated point sources + disk  &\vline  ~76356&\vline ~105&\vline ~152932  \\
\hline
\end{tabular}
\end{table*}

\begin{figure*}[ht]
\centering
\includegraphics[scale=0.8]{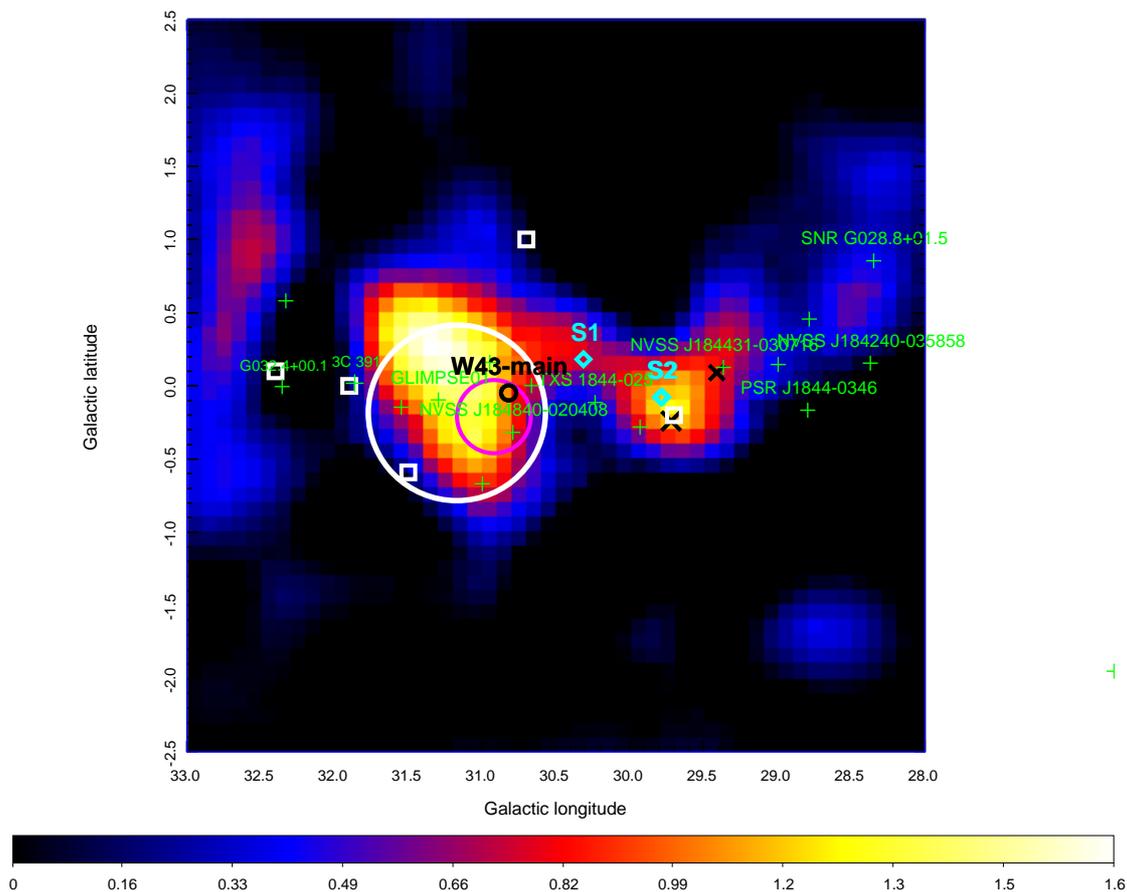}
\caption { The residual counts map above 10 GeV in the $5\deg \times 5\deg$ region near W43, with pixel size of $0.1\deg \times 0.1\deg$, smoothed with a Gaussian filter of $0.2^ \circ$. The color bars represent the counts per pixel. 
The white circle is the best-fit uniform disk template. The small black circle marks the position of the star forming region W43-main. The cyan diamonds are the weak sources S1 and S2.  The green crosses indicate the point sources listed in the 4FGL. The magenta circle marks the position and size of the TeV source HESS J1848-018. The two black crosses mark the position of the two HESS point sources, HESS J1846-029 (left) and HESS J1844-030 (right), respectively.   The white boxes are the positions of the nearby SNRs. 
}
\label{fig:resmap}
\end{figure*}

\subsection{Spatial analysis}
\label{sec:spatial_analy}
Firstly, we used the events above 10 GeV to study the spatial distribution of the \gray emissions. We note that there are four unassociated point sources (4FGL J1847.2-0141, 4FGL 1847.2-0200, 4FGL J1849.4-0117, and 4FGL 1850.2-0201) and two identified sources (4FGL 1848.7-0129 identified as GLIMPSE01 and 4FGL 1848.6-0202 identified as NVSS 184840-020408).   To study the excess \gray emission around W43, we excluded these four unassociated  4FGL sources  from our background model. 
After the likelihood fitting we subtracted the best-fit diffuse model and all the identified sources in the ROI; the resulting residual maps are shown in  Fig. \ref{fig:resmap}. We find  strong residuals toward  the direction of W43-main (marked with a black circle).

To study the morphology of the diffuse emission, we added a  disk on top of the model used in the likelihood analysis. We then varied the position and size of the disk to find the best-fit parameters. We compared the overall maximum likelihood of the model with $L$ (alternative hypothesis) and without  $L_{0}$ (null hypothesis) uniform disks, and defined the test-statistics (TS) of the disk model $-2({\rm ln}L_{0}-{\rm ln}L)$ following \citet{Lande12}. The best-fit result is a  disk centered at (ra=$282^{\circ}.22\pm 0.1$, dec=$-1^{\circ}.65\pm 0.1$) with $\sigma=0.6^{\circ}\pm 0.1$ and  a TS value of 270, corresponding to a significance of more than 16 $\sigma$.   We also tested whether or not this extended emission is composed of several independent point sources.  To do this we recovered the four unassociated point sources in the likelihood model. The -log(likelihood) function value is larger here than in the 2D disk template case, even with more free parameters.   To compare the goodness of the fit in the different models, we also calculated the Akaike information criterion (AIC) values for each model. The AIC was defined as $AIC = -2 \rm log(likelihood) +2 k$, where k is the number of free parameters in the model.  We list the the -log(likelihood) and corresponding AIC values in Table\ref{tab:loglike}.  We also note that the morphology of the residual hints at a deviation from a simple  disk; we tried to fit the residual with Gaussian and elliptical templates but found no significant improvement.  Thus in the following analysis we used the best-fit  disk as the spatial template.   %

%
%
The derived photon index above 10 GeV is $2.27 \pm 0.14$ and the total \gray flux can be estimated as $(6.1 \pm 0.7) \times 10^{-10} \ \rm ph.cm^-2.s^{-1}$ above 10 GeV, which reads $(2.8 \pm 0.4) \times 10^{35} \ \rm erg.s^{-1}$ above 1 GeV, if we assume a distance of 5.5 kpc \citep{Zhang2014} and a single power law spectrum. 
We find two other residuals to the west (right) of W43, which are marked as S1 and S2 in the residual map. Although adding point sources to the corresponding region does not significantly improve the likelihood  (the TS for S1 and S2 are 12 and 10, respectively), we included these two point sources in the following analysis. {There are three HESS TeV \gray sources in this region, which are HESS J1848-018, HESS J1846-029, and HESS J1844-030. The positions of the HESS sources are also marked in Fig.\ref{fig:resmap}.}

%

%
\subsection{Spectral analyses}
\label{sec:spectral_analy}
To further investigate the spectral property of the GeV emission toward W43 and the underlying particle spectra, we fixed the 0.6\deg\ uniform circle disk as the spatial model of the extended \gray emission and used a power law function to model the spectral shape. 

\begin{figure}[ht]
\centering
\includegraphics[width=1\hsize]{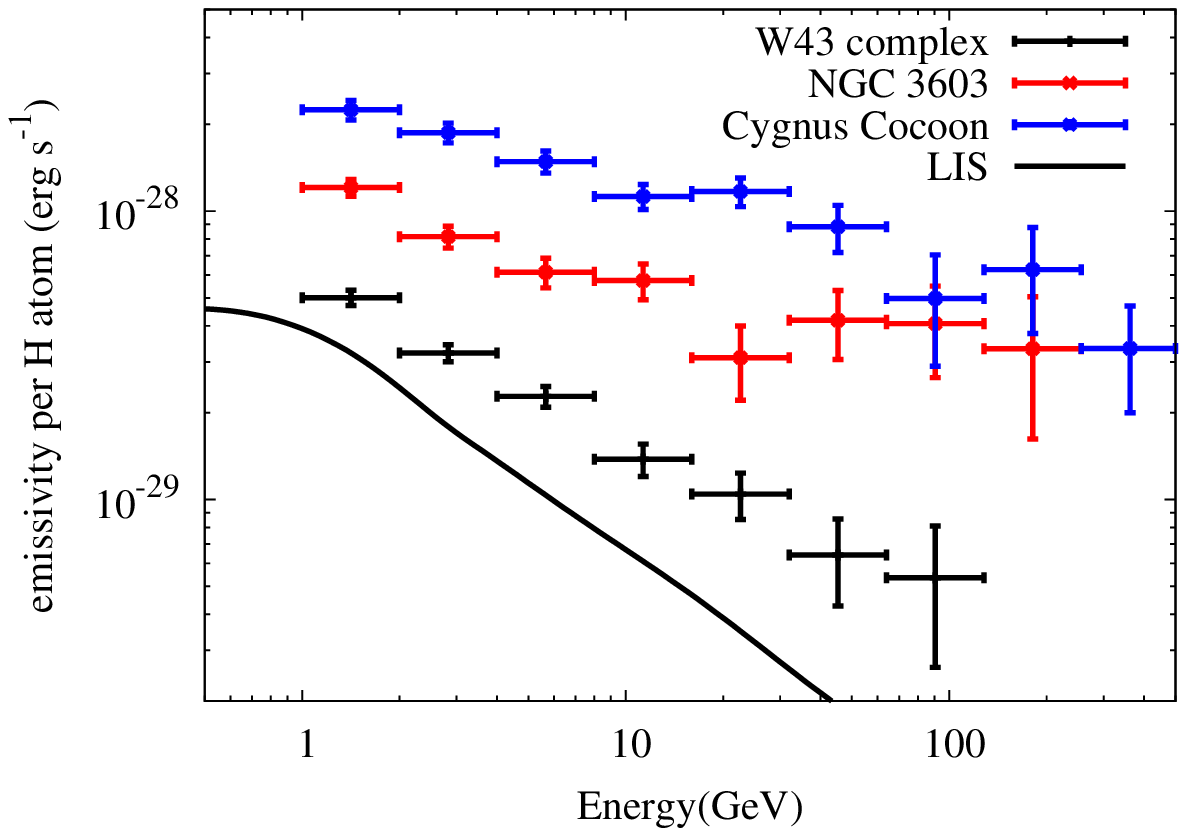}
\caption {
The SED of \gray emission toward W43 for a uniform disk spatial model with a radius of $0.6^{\circ}$, normalized to emissivity per H atom. The distance of 5.5 kpc  is used and the masses are derived in Sect. 3 . 
The solid curve represents the spectrum of  emissivity per H atom, assuming the energy distribution of protons is the same as the local intestellar spectrum (LIS) \citep{casandjian15}. Also plotted are the normalized SEDs of the Cygnus cocoon \citep{aharonian19} and NGC 3603 \citep{Yang17}. }
\label{fig:sed}
\end{figure}

\begin{figure*}[ht]
\centering
\includegraphics[scale=0.45]{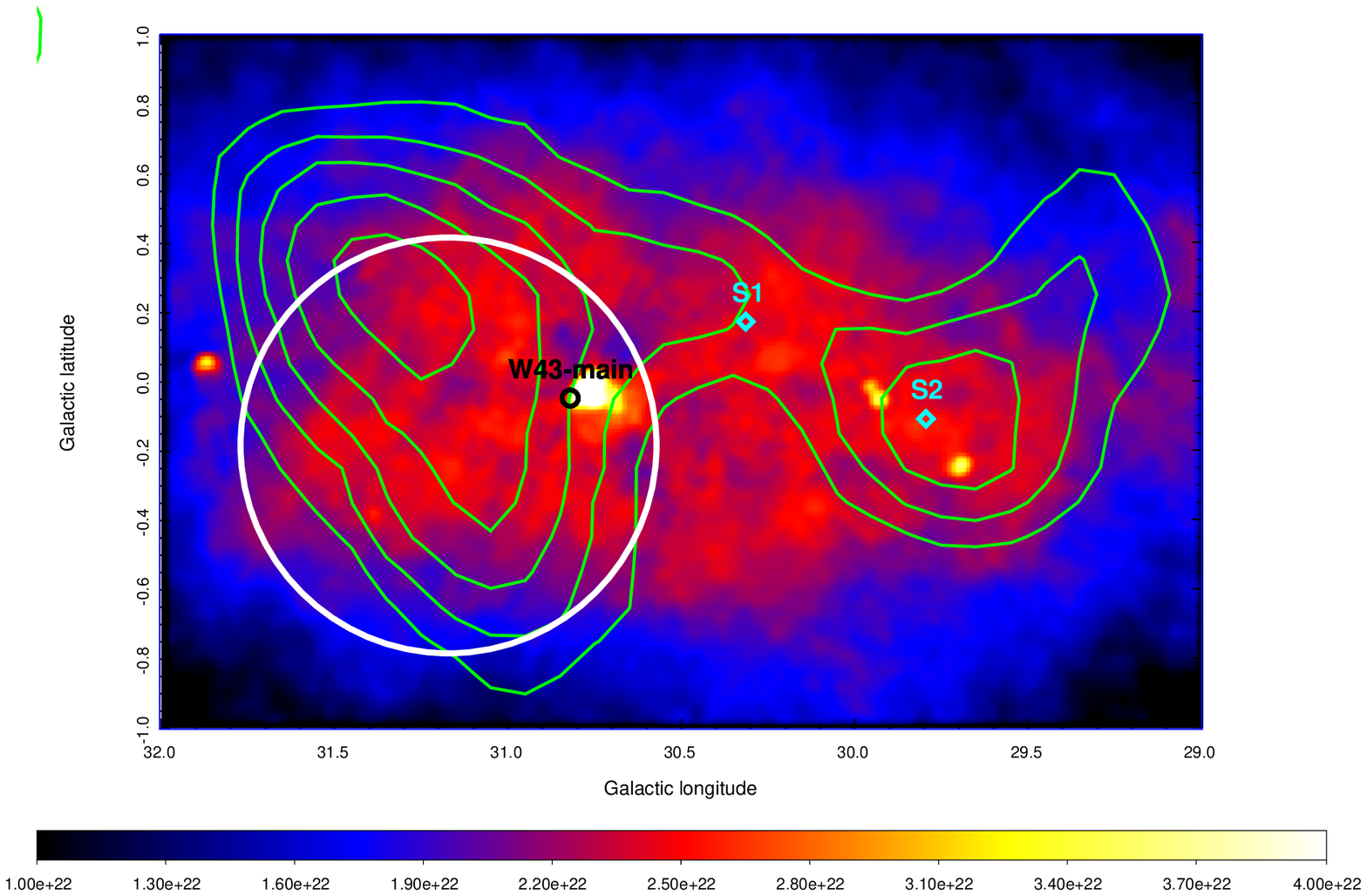}
\includegraphics[scale=0.45]{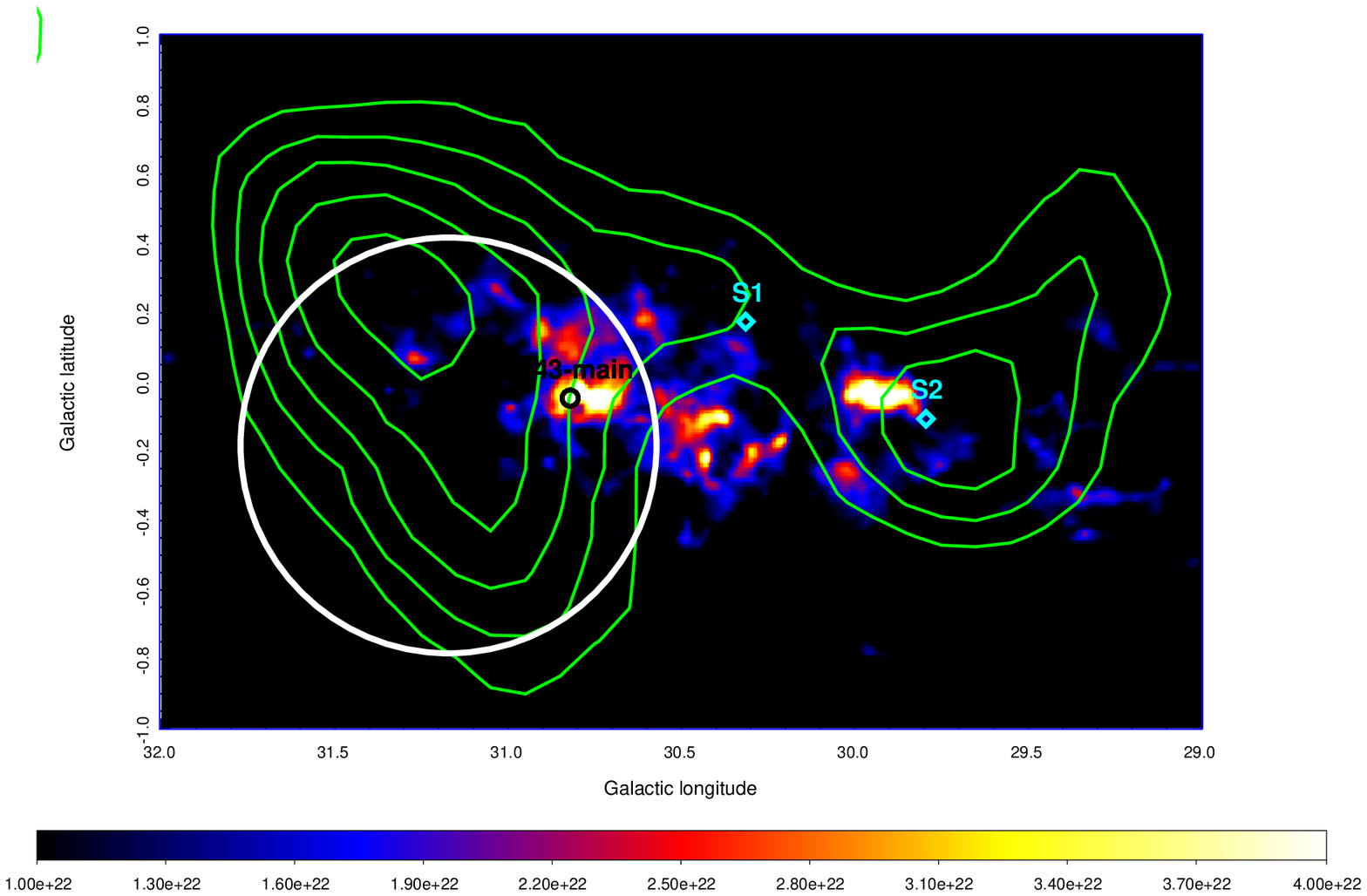}

\includegraphics[scale=0.45]{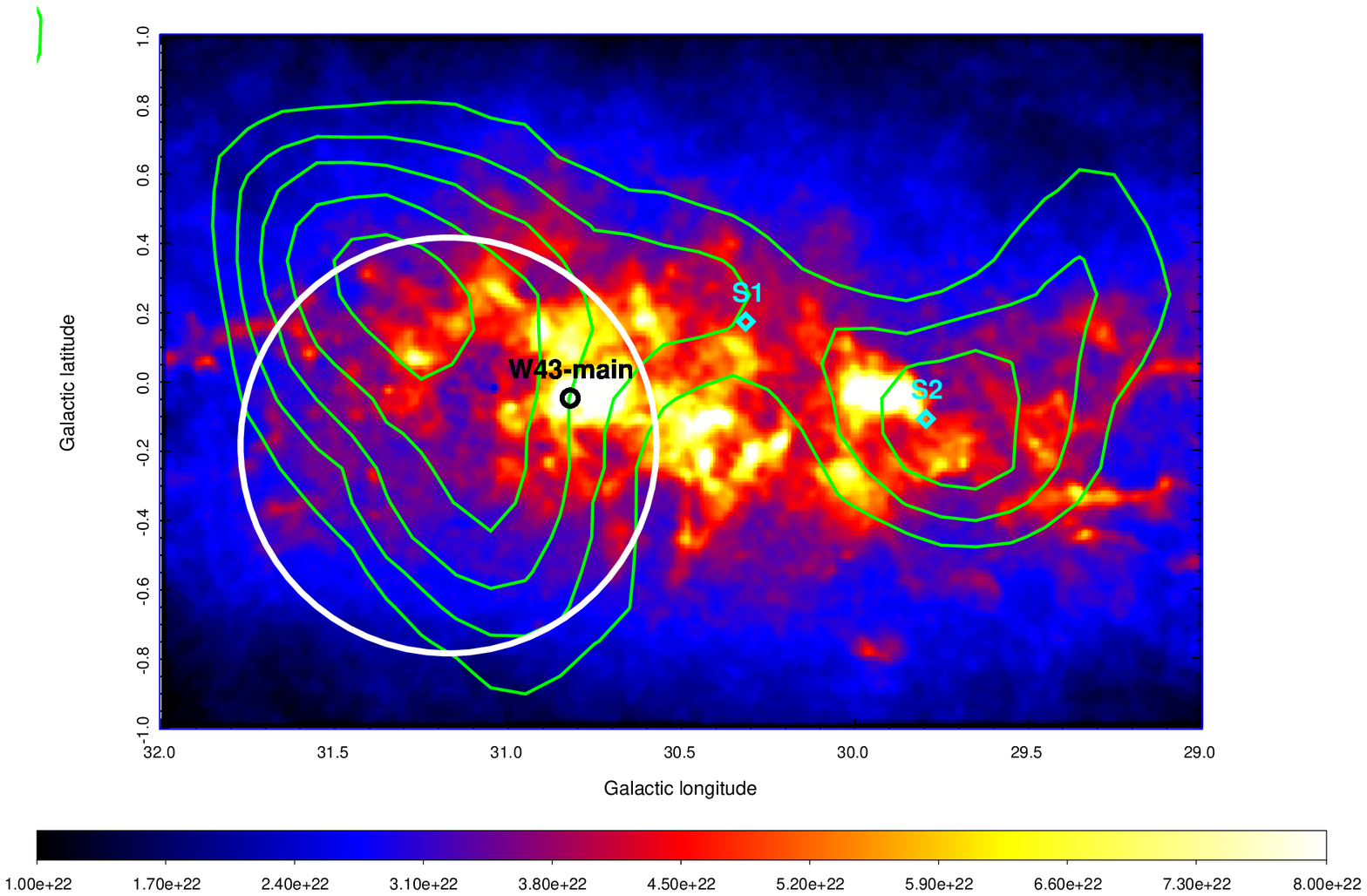}
\caption { 
Maps of the \ion{H}{i} (top-left), H$_2$ (top-right), and total (bottom) gas columns.  
The labels are the same as in \figurename~\ref{fig:resmap}. The green contours represent the \gray emission, the contour levels are the counts per pixel of 0.5, 0.75, 1.0, 1.25, and 1.5 in the residual map above 10 GeV (figure\ref{fig:resmap}). The color bars represent the gas column density in $\rm cm^{-2}$. For details, see the context in Sect.~\ref{sec:Gas}.
}
\label{fig:gasdis}
\end{figure*}

We divided the energy range 1-300 GeV into nine logarithmically spaced energy bins and derived the spectral energy distribution (SED) via the maximum likelihood analysis in each energy bin. The results are shown in Fig. \ref{fig:sed}.
The significance of the signal detection for each energy bin exceeds $2\sigma$.
We calculated 68\% statistical errors for the energy flux densities.  We further divided the \gray flux by the average gas column density (see Sect. \ref{sec:Gas} for details) to get the \gray emissivity per H atom. Assuming the \grays are produced by an interaction between CRs and ambient gas, the \gray emissivity per H atom should be proportional to the CR density. We performed the same procedure for the \gray SEDs in NGC 3603 \citep{Yang17} and the Cygnus cocoon \citep{aharonian19}; the results are also shown in Fig. \ref{fig:sed} for comparison. 
In addition, we estimated the uncertainty caused by the imperfection of the Galactic diffuse background model by artificially changing the normalization  by $\pm$6\% from the best-fit value for the six low energy bins, and considered the maximum flux deviation of the source as the systematic error \citep{Abdo09}.  

%

%
%

%

\section{Gas content around W43}
\label{sec:Gas}
With the \ion{H}{i} data from the \ion{H}{i}/OH/Recombination line survey of the inner Milky Way (THOR, \citealt{beuther2016, Wang2019}), we estimated the \ion{H}{i} column density using \citep[e.g., ][]{wilson2013}:
\begin{equation}
N_{\rm H} = 1.8224\times10^{18}\: \int T_{\rm S}(v) \tau(v)\, dv.
\label{eq:coldens_hi}
\end{equation}
The optical depth corrected spin temperature is $T_{\rm S}(v)=T_{\rm B}(v)/(1-{\rm e}^{-\tau(v)})$, where $T_{\rm B}$ is the brightness temperature of the \ion{H}{i} emission. The optical depth data from \citet{Wang2019} is used to correct for the spin temperature channel by channel. We further followed the method described in \citet{Bihr2015} and corrected the column density for diffuse continuum absorption using the THOR+VGPS\footnote{The Very Large Array Galactic Plane Survey, \citealt{stil2006}} 1.4~GHz continuum data \citep[C+D+single dish GBT, see ][]{Wang2018}. The derived \ion{H}{i} column map integrated in the velocity range $v_{\rm LSR}$=60--120~km~s$^{-1}$ is shown in the top-left panel of Fig.~\ref{fig:gasdis}.

The $^{13}$CO(1--0) and $^{12}$CO(1--0) data used to characterize the molecular gas content properties are from the Galactic Ring Survey \citep[GRS, ][]{Jackson2006} and the FOREST unbiased Galactic plane imaging survey conducted with the Nobeyama 45~m telescope \citep[FUGIN, ][]{Umemoto2017}, respectively. Assuming  optically thin emission, we estimated the column density of the $^{13}$CO molecule with the equation \citep{wilson2013}:
\begin{equation}
N(^{13}{\rm{CO}})  = 3.0\times10^{14} \: \frac{\int T_{\rm{MB}}(v)\, dv}{1-{\rm e}^{-5.3/T_{\rm{ex}}}},
\label{eq:coldens_co}
\end{equation}
where $N(^{13}\rm{CO})$ is the column density of the $^{13}$CO molecule in cm$^{-2}$, $dv$ is the velocity in km\,s$^{-1}$, $T_{\rm{MB}}$ is the main beam brightness temperature in K, and $T_{\rm{ex}}$ is the excitation temperature in K. Assuming that $^{12}$CO and $^{13}$CO share the same excitation temperatures and the $^{12}$CO line is optically thick, we can estimate $T_{\rm{ex}}$ following the formula \citep{wilson2013}:
\begin{equation}
T_{\rm{ex}}=\frac{5.5}{{\rm ln}\left(1+\frac{5.5}{T_{\textrm{mb}}(^{12}\rm{CO})+0.82}\right)},
\label{eq:tex}
\end{equation}
where $T_{\rm{mb}}(^{12}\rm{CO})$ is the main-beam brightness temperature of the  $^{12}\rm{CO}$(1--0) line. The $T_{\rm{ex}}$ were calculated for regions where $T_{\rm{mb}}(^{12}\rm{CO})$ is above the 5$\sigma$ level (2~K), which results in a $T_{\rm{ex}}$ between $\sim$5 and 45~K. For regions where $T_{\rm{mb}}(^{12}\rm{CO})$ is below the 5$\sigma$ level (2~K), an upper limit of 5~K for $T_{\rm{ex}}$ is applied. For the galactocentric distance of 4.6~kpc of W43, the fractional abundance of $^{13}$CO relative to H$_2$ is estimated to be 4.3$\times10^{-6}$ following the relations reported by \citet{Giannetti2014}. We converted $N(^{13}\rm{CO})$ to $N(\rm{H_2})$ with this abundance.

 Since $^{13}$CO does not trace the diffuse molecular gas \citep{Roman-Duval2016}, we used the FUGIN $^{12}$CO data \citep{Umemoto2017} to derive the column density map of the diffuse gas. Assuming optically thin emission, we followed the standard method described in \citet{Feng2016} and derived an H$_2$ column density map from $^{12}\rm{CO}$ with the same excitation temperature and $^{12}$CO abundance mentioned earlier.  We then used the H2 column densities derived from $^{12}$CO instead of the column densities derived from $^{13}$CO, where the latter is 3$\sigma$ lower than the former. The combined $N(\rm{H_2})$ map ($v_{\rm LSR}$=60--120~km~s$^{-1}$) is shown in the top-right panel of Fig.~\ref{fig:gasdis}. Compared to the $N(\rm{H_2})$ map derived with only $^{13}$CO emission, the combined $N(\rm{H_2})$ map has a molecular gas mass that is 56\% larger.

We smoothed both the $N(\rm{H})$ and $N(\rm{H_2})$ column density map to the same spatial resolution (46\arcsec) and combined them to make the total gas column density map in units of hydrogen atoms cm$^{-2}$ shown in the bottom panel of Fig.~\ref{fig:gasdis}. 

The total gas mass in the \gray emission region  is $3.5\times 10^6 M_{\odot}$. Assuming spherical geometry of the \gray emission region, its radius is estimated as $r = d \times \theta \sim 5500\ {\rm pc} \times (0.6\deg \times \pi/180\deg)\ {\rm rad} \sim 60\ {\rm pc}$. Thus the average gas number density over the volume is about $ 140\ cm^{-3}$.

%
%

%

%
%

%

%

%
\section{Discussions}
\label{sec:discussion}
\subsection{Different sources in the region}
 There are three pulsars in the disk region from the ANTF pulsar catalog \footnote{https://www.atnf.csiro.au} \citep{atnf}. They are PSR J1847-0130, PSR J1848-0055, and PSR B1845-01. Their spin-down luminosities are $1.7\times 10^{32} $, $2.6\times 10^{33} $, and $7.2\times 10^{32} \rm ergs/s$, while their distances are 5.8, 7.4, and 4.4 kpc, respectively.  We cannot rule out the possibility that these extended \gray emissions are indeed PWNs, but there are no pulsars in this region that have enough spin-down power to provide the $3\times10^{35} ~\rm erg/s$ \gray luminosity, which makes this scenario quite unlikely. 

There are also several SNRs near this region, including SNR G31.5-0.6 which is located inside the \gray emission region. The SNR G31.5-0.6 is an SNR with an incomplete shell and whose distance was estimated to be 12.9 kpc by using the $\Sigma-$D relation \citep{case98}. 

The mixed-morphology SNR 3C 391 is also to the east of the \gray emission region. It is a mid-aged SNR with an age of about 4000 years and a distance of 7.2 kpc. The GeV emissions from this source are studied in detail in \citet{ergin14}. 
It is possible that the extended \grays in the W43 region are produced by the interaction of ambient gas with the CRs that escaped from these SNRs. 

Furthermore, the star forming region W43 is another natural acceleration site of the CRs. Indeed, the spectral shape and spatial extension are similar to those measured in other massive star clusters, such as Westerlund 2 \citep{Yang18}, NGC 3603 \citep{Yang17}, and the Cygnus cocoon \citep{Ackermann11,aharonian19}. The center of W43-main also harbors a Wolf-Rayet/O-star cluster \citep{blum99}. Thus  the CRs accelerated by this cluster interacting with ambient gas provides a natural explanation of the extended \gray emissions.

\subsection{Origin of the extended \gray emissions}
As calculated in the last section, the average number density of the target protons is  $\sim \rm 140\ cm^{-3}$.  In such dense regions, the pion-decay \grays will dominate the leptonic \grays if we assume a reasonable e/p ratio \citep{gabici07}. The leptonic origin of \grays is still feasible in the PWN scenario but, as mentioned above, it is quite unlikely  due to the limited power of the pulsars in the vicinity. Thus in the discussion below we assume the \grays are produced in the pion-decay process from the interaction of CRs with ambient gas. To provide the \gray luminosity of $(2.8 \pm 0.4) \times 10^{35} \ \rm erg.s^{-1}$, the required total CR energy would be $\sim2.3 \pm 0.3 \times 10^{48} \ \rm erg$ (above 10 GeV), which is orders of magnitude smaller than those in other similar systems \citep{aharonian19}. However, the derived CR density is still significantly larger than densities measured in the solar neighborhood. In Fig. \ref{fig:sed} we plot the predicted \gray flux assuming the CR density in W43 is the same as the local measurement.

\begin{figure}[ht]
\centering
\includegraphics[width=1\hsize]{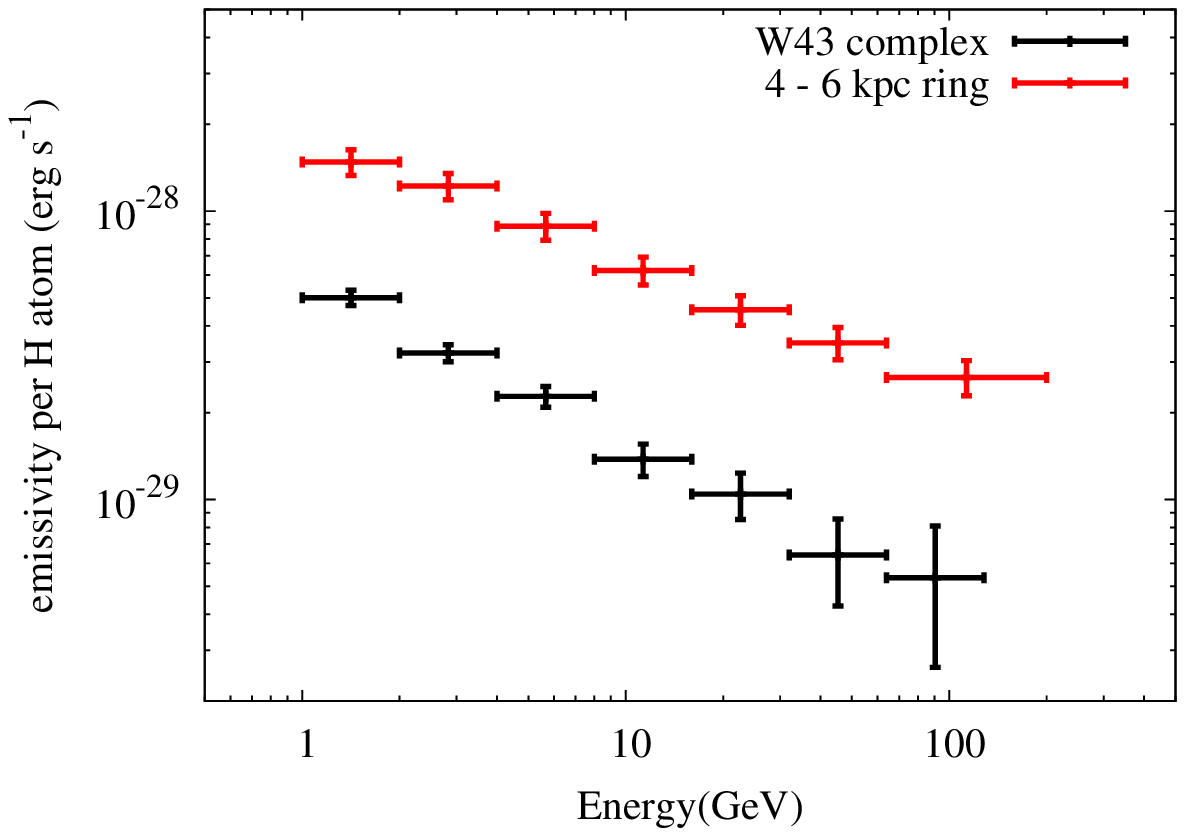}
\caption {The \gray SEDs of W43-main and in the 4-6 kpc ring from \citet{yang16}. 
}
\label{fig:sed_cp}
\end{figure}

 As calculated in \citet{yang16} and \citet{fermi_diffuse}, the CR densities  do  reveal an enhancement and spectral hardening in the inner Galaxy. We thus compared the \gray emission we derived here with the average \gray spectrum in the 4-6 kpc ring in our Galaxy (the data points are from \citet{yang16}).   The results are shown in Fig. \ref{fig:sed_cp}. The \gray spectrum from W43 is similar to that derived for the 4-6 kpc rings, but the total normalization is 40\% smaller. Indeed, as mentioned in \citet{aharonian20}, the enhancement of CR density in the 4-6 kpc ring is not the global variation of the level of the CR sea. Instead, the enhancement is caused by the fact that most of the active star-forming regions and, therefore, the potential particle accelerators, are located within the 4-6 kpc ring. These accelerators can create the CR-enhanced region in their vicinity. Since the CR density in such  regions  depends on the strength and the age of the accelerator,  the densities can differ from one region to another. Thus it is not surprising that the CR density in W43 is lower than the average for the 4-6 kpc ring.

As mentioned in \citet{aharonian19}, the CR density distribution around several massive star clusters reveals a unique 1/r profile. We perform a similar analysis for W43 to derive the CR spatial distribution.

\begin{figure*}[ht]
\centering
\includegraphics[scale=0.8]{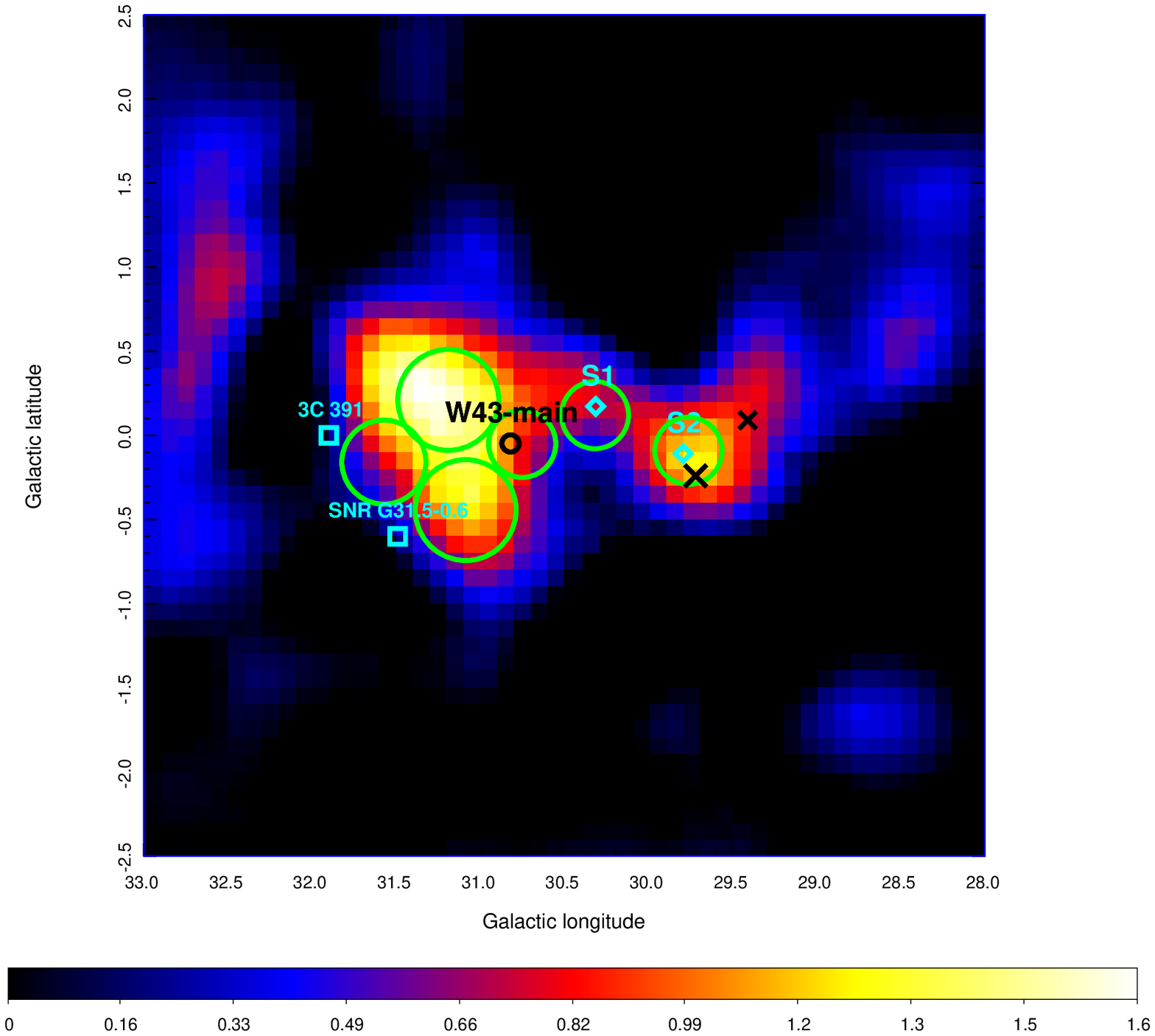}
\caption {The same as Fig. \ref{fig:resmap} but overlaid with the disk region we used to extract the CR profile. 
}
\label{fig:resseg}
\end{figure*}

\begin{figure*}[ht]
\centering
\includegraphics[scale=0.4]{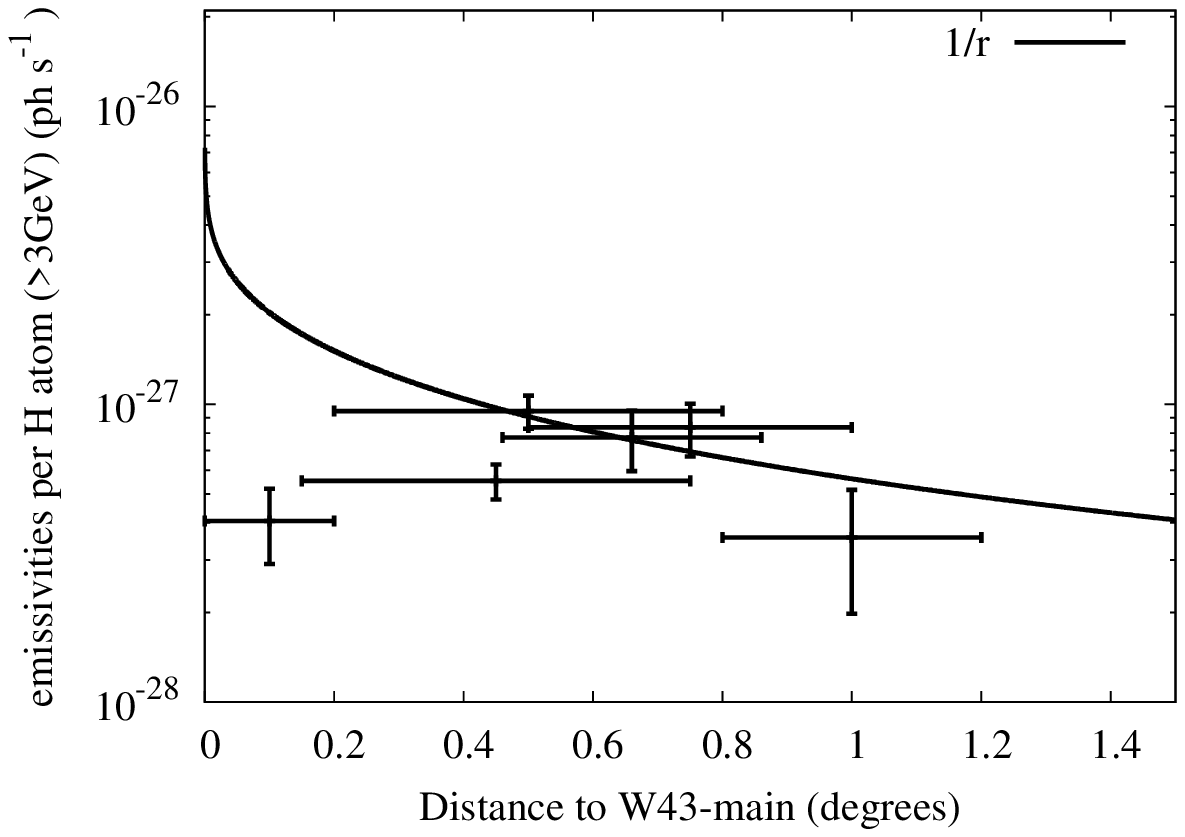}
\includegraphics[scale=0.4]{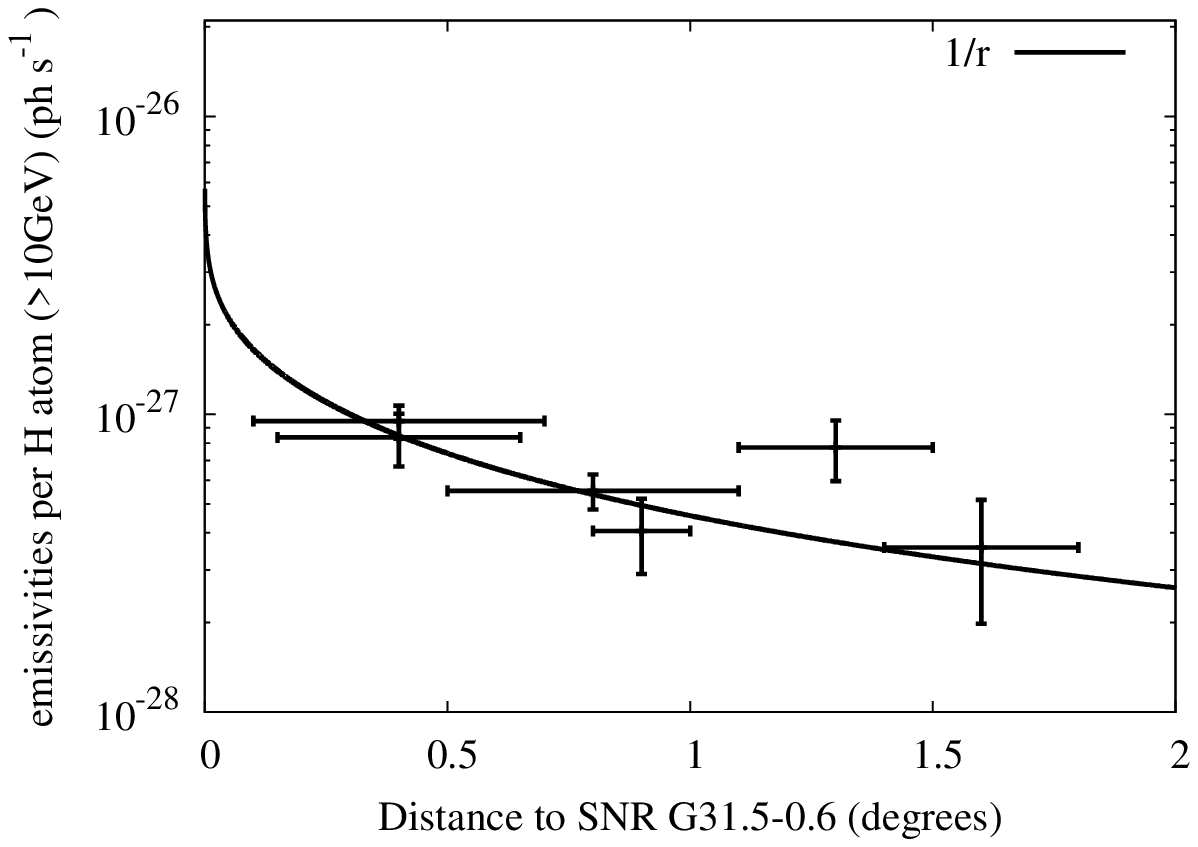}
\includegraphics[scale=0.4]{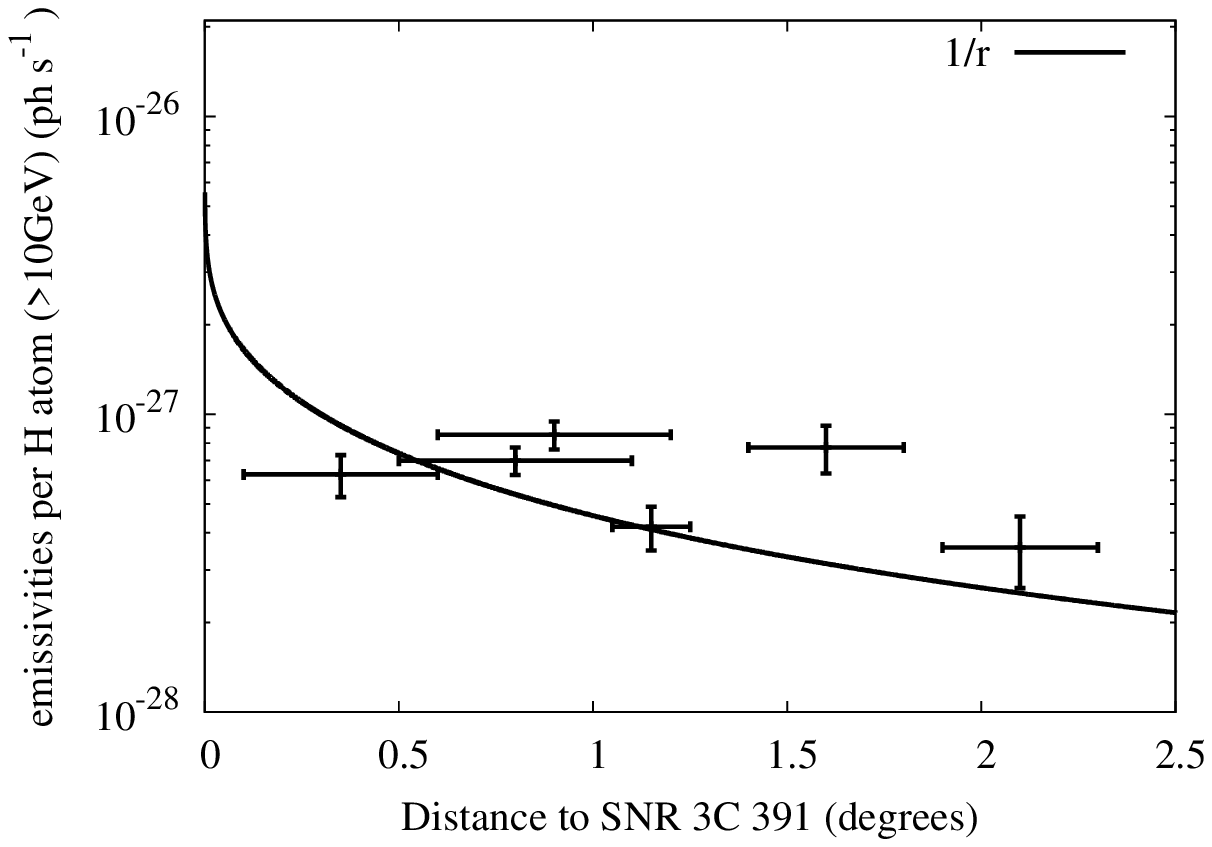}

\caption {The \gray emissivity per H atom with respect to W43-main (left panel), SNR G31.5-0.6 (middle panel), and SNR 3C391 (right panel). The curves are the projected 1/r profile. 
}
\label{fig:pro}
\end{figure*}

To do this, rather than divide the \gray emission region into rings as in \citet{aharonian19}, we divided it into four disk templates, as shown in Fig. \ref{fig:resseg}. This is because the potential accelerator, the massive star cluster inside W43-main, is far from the center of the \gray emission region. We also derived  the CR density in the two weak sources S1 and S2.  The derived photon indices between 10~GeV and 300~GeV are $2.0 \pm 0.2, 2.4 \pm 0.3, 2.2 \pm 0.2,$ and $2.3\pm0.2$. We found no significant spectral change in any of these regions, which further supports the idea that the extended emissions are of the same origin. We then performed a likelihood analysis and derived the \gray flux in each disk template. We then divided the \gray flux by the total hydrogen atom numbers derived from the mass to get the \gray emissivity per H atom, which is proportional to CR density. The derived CR profile is shown  in the left-hand panel of Fig. \ref{fig:pro}.  We also performed a jackknife test by slightly changing the position and size of these disk templates (by 0.1 degree). We found the results to be stable and the resulted differences are included in the error bars of the fluxes.  We found no 1/r dependence in this case and the CR density near W43-main is significantly lower than in other parts. One  explanation for such a profile is that the CRs are not accelerated exclusively in the star cluster in W43-main. Other sources, such as SNRs and other massive star clusters, reside in the W43 complex and may be also responsible for the acceleration of the CRs. We cannot exclude the possibility that these potential accelerators have not been observed in other wavebands, especially considering the very high column density and extinction toward W43. The massive stars may be invisible in the optical band due to high extinction. Ongoing near infrared surveys such as APOGEE \citep{apogee1} have the ability to find these obscured massive stars in dense clouds, which may play an important role in determining the origin of CRs in such a system.  If there are multiple sources, no CR spatial distribution can be ruled out. But it is also possible that the CRs are dominantly contributed by another single source.  To check this possibility, we also plotted the CR distribution with respect to SNR G31.5-0.6 and SNR 3C 391, which are located to the lower-left (southeast) and left (east)  of the \gray emission regions (see Fig. \ref{fig:resseg}).  The profiles of the \gray emissivities are also shown in Fig. \ref{fig:pro}.  In the case of 3C 391, we find no dependence, while in the case of SNR G31.5-0.6 the distribution reveals a clear 1/r profile. The $\chi^2/d.o.f$ for the 1/r profile is 5.6/5, while for a uniform profile it is 16.5/5.  We note that  the S2 position coincides with SNR G29.7-0.3/HESS J1846-029/PSR J1846-0258. It is possible that S2 is the low energy counterpart of the TeV source HESS J1846-029.   Thus we also estimated the profile by ignoring S1 and S2. In this case, the $\chi^2/d.o.f$ is 2.5/3 for a 1/r profile, and 13.3/3 for a uniform profile. The distance of SNR G31.5-0.6 was estimated to be 12.9 kpc, using the $\Sigma-$D relation \citep{case98}. If this is true, SNR G31.5-0.6 can hardly be responsible for the CRs illuminating W43 region. But it is also possible that SNR G31.5-0.6 does not obey the $\Sigma-$D relation used in deriving the distance.  In this case, SNR G31.5-0.6 can also be regarded as a potential accelerator of CRs in this region.

\subsection{Acceleration mechanisms}\label{sec:conc}

\begin{table*}
\centering
\caption{Physical parameters of  three extended $\gamma$-ray structures and the star forming region. }
\begin{tabular}{|c|ccc|}
 \hline
Source&  Cyg Cocoon & NGC 3603 & W43 \\
 \hline
 \hline
Extension (pc) &  50 & 130 & 60\\
L$_{kin}$ of cluster (erg/s) & $(3\pm 1) \times$10$^{38}$   & $(2\pm 1)\times$10$^{38}$ & $(3\pm 1)\times$10$^{37}$\\
SFR ($M_{\odot}/yr$) &  0.06 & 0.002 & 0.32 \\
$W_p(> 10 \rm GeV)$ (erg) & $(1.0\pm0.2) \times 10^{49}$    & $(7.5\pm 2.5) \times 10^{49}$ & $(2.3\pm 0.3) \times 10^{48}$ \\
Normalized  $W_p(> 10 \rm GeV)$ (erg) & $(6.7\pm1.2) \times 10^{49}$    & $(7.5\pm 2.5) \times 10^{49}$ & $(1.2\pm 0.2) \times 10^{49}$ \\
\hline
\end{tabular}

\tablefoot{The wind kinematic power of Cygnus OB2 is from \citet{Ackermann11}; the discussion of the wind powers of NGC 3603 and W43 are in the text. The SFRs of W43 and the Cygnus X region can be found in \citet{luong16}, and for NGC3603 in \citet{beccari10}. The total CR energy $W_p$ is derived from the \gray emission region. The normalized $W_p$ is derived assuming the CRs injected by Cygnus OB2 and by W43 occupy the same volume as that of NGC 3603, see the text for details.     }
\label{tab:sf}
\end{table*}

Recently, \citet{aharonian19} proposed that young star clusters can be an alternative source population of  Galactic CRs, and that the \gray emissions around such objects can be a powerful tool to diagnose the acceleration of CRs  and the propagation of CRs in the vicinity of sources.  Several such systems have already been detected in the \gray band, such as  the Cygnus cocoon \citep{Ackermann11, aharonian19}, Westerlund 1 \citep{Abramowski12}, Westerlund 2 \citep{Yang18}, NGC 3603 \citep{Yang17}, and 30 Dor C \citep{Abramowski15}. Here, we report a statistically significant detection of an extended \gray signal from the direction of another  young star forming region, W43.   Like in the other systems, the spectrum of this source is harder than the local CR spectra. We argue that the most likely origin of the detected emission is the interactions of CRs with the surrounding  gas, which are accelerated  in the  star forming region.

However, the exact CR accelerator is not clear at this moment. The most natural source, the massive star cluster in W43-main, is likely not the source according to the CR density profile. A source located to the southeast of W43 can recover a 1/r profile observed in other similar systems. We note that the 1/r profile implies a continuous injection of CRs.  Thus the size of the \gray emission region can be used to constrain the age of the accelerator. If we attribute the farthest \gray emission S2 to be related to the source, the size l of the \gray emission region is larger than 150 pc. Thus the age should be larger than  $T=l^2/2D$. If we take  the standard diffusion coefficient in the Galactic plane of $10^{28} ~\rm cm^2/s $ into account, the derived age is larger than $10^5$ years, which seems too long for an SNR and also favors a massive star cluster scenario of CR accelerations.  We note that it is quite common for OB stars to be missed in such a dense region; the APOGEE survey has revealed four times more OB stars than identified before in the W345 complex \citep{apogee2}. Such dedicated near infrared surveys will shed light on the origin of these diffuse \gray emissions and play an important role in the study of the CR origin in our Galaxy. 

Another interesting issue is that the \gray luminosity is proportional to the SFR in starburst galaxies \citep{fermi_galaxy}. The natural explanation is that those galaxies are CR calorimeters and the \gray luminosity is the same as the CR injection rate. Thus the aforementioned relationship indicates that the CR injection rate depends on the SFR. 

To extend the  study to nearby Galactic systems,  we included another two \gray detected Galactic star forming (starburst) regions: the  Cygnus cocoon (powered by Cygnus OB2) \citep{Ackermann11, aharonian19} and NGC 3603 \citep{Yang17}. However, due to the compact size, most of the accelerated CRs will escape the source region before losing their energies. Thus, rather than compare the \gray luminosities, we calculated the total CR energies from the \gray observations and the gas distributions. As derived in Sect. 3, the total CR energy in W43 is $(2.3\pm 0.3) \times 10^{48} \rm erg$. The total CR energy in NGC3603 and the Cygnus cocoon are $(7.5\pm 2.5) \times 10^{49} \rm erg$ \citep{Yang17} and $(1.0\pm 0.2) \times 10^{49} \rm erg$ \citep{aharonian19}, respectively. However, as mentioned in \citet{aharonian19}, the total CR energy estimated from \gray emission may be biased due to the fact that the CRs may occupy a much larger volume but cannot be traced by \grays due to the low gas density therein.  Of these three cases, the \gray emission region of NGC3603 is much more extended than the other two. Thus, to make the comparison feasible, we normalized the CR density to the same volume by assuming a 1/r distribution of CR density, although we did not find such a relationship in W43. 

\citet{luong16} derived the the immediate past SFR from the radio continuum for the Cygnus X region (which includes Cygnus OB2) and W43. We estimated the average SFR  of NGC 3603 by simply dividing  the mass of a cluster by its age \citep{beccari10}. We find that although W43 has much larger SFRs than the other two, the injected CR energy is much lower, which is contrary to the relationship found in galaxies. 

We further investigated the relationship between CR energy and wind power in these regions. As estimated in \citet{santamar06}, the wind power of a single Wolf-Rayet star or early type-O star can be as large as $10^{37} ~\rm erg/s$, while for B-stars the value drops by orders of magnitude. Thus, the wind power in these regions should be dominated by O-stars and Wolf-Rayet stars.    Cygnus OB2 and NGC 3603 both contain several dozen O-stars. \citet{Ackermann11}  estimated the total wind power of Cygnus OB2 to be $2-3 \times 10^{38} ~\rm erg/s$. \citet{wright10} mentioned that the number of O-stars in NGC3603  is about two-thirds of that in Cygnus OB2. Thus we simply estimated the total wind power in NGC3603 to also be two-thirds of that in Cygnus OB2. We note that this estimation is quite conservative since although there are fewer O-stars in NGC3603, four of them have already been  identified as Wolf-Rayet stars \citep{harayama08}. For W43, so far only one Wolf-Rayet and two O-stars have been identified \citep{blum99}. Thus we estimated the wind power to be about $3 \times 10^{37}~\rm erg/s$. The physical properties of all three regions are listed in Table \ref{tab:sf}.  In this case we found a clear correlation between the wind power and total CR energy. We regarded it as an indication that in such systems CRs are indeed accelerated by star winds. 

\section{Conclusion}

    In this paper we analyze the  \gray emission from the Galactic mini-starburst W43. We find an extended \gray emission with a radius of $0.6^{\circ}$ and a significance of $16~\sigma$. The emission reveals a hard spectrum, with a spectral index of about 2.3. The spatial and spectral features are similar to the \gray emissions in other massive star clusters such as NGC 3603 \citep{Yang17}, Westerlund 2  \citep{Yang18}, and the Cygnus cocoon \citep{aharonian19}. Due to the high gas density in this region, the most probable \gray emission mechanism is the pion-decay process that is the result of the interaction of CR protons with the ambient gas. In comparison with similar systems, we find that the total CR energy in the star forming region is  better correlated with the wind power than with the SFR. We derived the CR profile in the \gray emission region. We find that, unlike the Cygnus cocoon and Westerlund 2, the CR distribution does not reveal a 1/r profile over the central massive star clusters.  However, it is quite likely that the massive star  associations/clusters were missed  in former studies due to high extinction in this dense region. Dedicated near infrared surveys such as APOGEE \citep{apogee1} will shed light on the origin of these diffuse \gray emissions and play an important role in the study of the CR origin in our Galaxy.


\section*{Acknowledgements}

Ruizhi Yang is supported by  the NSFC under grants 11421303 and  the national youth thousand talents program in China.
\bibliographystyle{aa}
\bibliography{w43}

\begin{thebibliography}{45}
\expandafter\ifx\csname natexlab\endcsname\relax\def\natexlab#1{#1}\fi

\bibitem[{{Abdo} {et~al.}(2009){Abdo}, {Ackermann}, {Ajello}, {Baldini},
  {Ballet}, {Barbiellini}, {Baring}, {Bastieri}, {Baughman}, {Bechtol},
  {Bellazzini}, {Berenji}, {Blandford}, {Bloom}, {Bonamente}, {Borgland},
  {Bouvier}, {Bregeon}, {Brez}, {Brigida}, {Bruel}, {Burnett}, {Buson},
  {Caliandro}, {Cameron}, {Caraveo}, {Casandjian}, {Cecchi}, {{\c C}elik},
  {Chekhtman}, {Cheung}, {Chiang}, {Ciprini}, {Claus}, {Cohen-Tanugi},
  {Cominsky}, {Conrad}, {Cutini}, {Dermer}, {de Angelis}, {de Palma}, {Digel},
  {Dormody}, {Silva}, {Drell}, {Dubois}, {Dumora}, {Farnier}, {Favuzzi},
  {Fegan}, {Focke}, {Fortin}, {Frailis}, {Fukazawa}, {Funk}, {Fusco},
  {Gargano}, {Gasparrini}, {Gehrels}, {Germani}, {Giavitto}, {Giebels},
  {Giglietto}, {Giordano}, {Glanzman}, {Godfrey}, {Grenier}, {Grondin},
  {Grove}, {Guillemot}, {Guiriec}, {Hanabata}, {Harding}, {Hayashida}, {Hays},
  {Hughes}, {Jackson}, {J{\'o}hannesson}, {Johnson}, {Johnson}, {Johnson},
  {Kamae}, {Katagiri}, {Kataoka}, {Katsuta}, {Kawai}, {Kerr}, {Kn{\"o}dlseder},
  {Kocian}, {Kuss}, {Lande}, {Latronico}, {Lemoine-Goumard}, {Longo},
  {Loparco}, {Lott}, {Lovellette}, {Lubrano}, {Makeev}, {Mazziotta}, {McEnery},
  {Meurer}, {Michelson}, {Mitthumsiri}, {Mizuno}, {Moiseev}, {Monte},
  {Monzani}, {Morselli}, {Moskalenko}, {Murgia}, {Nakamori}, {Nolan}, {Norris},
  {Nuss}, {Ohsugi}, {Okumura}, {Omodei}, {Orlando}, {Ormes}, {Paneque},
  {Parent}, {Pelassa}, {Pepe}, {Pesce-Rollins}, {Piron}, {Porter}, {Rain{\`o}},
  {Rando}, {Razzano}, {Reimer}, {Reimer}, {Reposeur}, {Ritz}, {Rodriguez},
  {Romani}, {Roth}, {Ryde}, {Sadrozinski}, {Sanchez}, {Sander}, {Saz
  Parkinson}, {Scargle}, {Schalk}, {Sgr{\`o}}, {Siskind}, {Smith}, {Smith},
  {Spandre}, {Spinelli}, {Strickman}, {Suson}, {Tajima}, {Takahashi},
  {Takahashi}, {Tanaka}, {Thayer}, {Thayer}, {Thompson}, {Tibaldo}, {Tibolla},
  {Torres}, {Tosti}, {Tramacere}, {Uchiyama}, {Usher}, {Vasileiou}, {Venter},
  {Vilchez}, {Vitale}, {Waite}, {Wang}, {Winer}, {Wood}, {Yamazaki}, {Ylinen},
  \& {Ziegler}}]{Abdo09}
{Abdo}, A.~A., {Ackermann}, M., {Ajello}, M., {et~al.} 2009, \apjl, 706, L1

\bibitem[{{Abramowski} {et~al.}(2012){Abramowski}, {Acero}, {Aharonian},
  {Akhperjanian}, {Anton}, {Balzer}, {Barnacka}, {Barres de Almeida},
  {Becherini}, {Becker}, {Behera}, {Bernl{\"o}hr}, {Birsin}, {Biteau},
  {Bochow}, {Boisson}, {Bolmont}, {Bordas}, {Brucker}, {Brun}, {Brun}, {Bulik},
  {B{\"u}sching}, {Carrigan}, {Casanova}, {Cerruti}, {Chadwick}, {Charbonnier},
  {Chaves}, {Cheesebrough}, {Chounet}, {Clapson}, {Coignet}, {Cologna},
  {Conrad}, {Dalton}, {Daniel}, {Davids}, {Degrange}, {Deil}, {Dickinson},
  {Djannati-Ata{\"i}}, {Domainko}, {O'C.~Drury}, {Dubois}, {Dubus}, {Dutson},
  {Dyks}, {Dyrda}, {Egberts}, {Eger}, {Espigat}, {Fallon}, {Farnier}, {Fegan},
  {Feinstein}, {Fernandes}, {Fiasson}, {Fontaine}, {F{\"o}rster},
  {F{\"u}{\ss}ling}, {Gallant}, {Gast}, {G{\'e}rard}, {Gerbig}, {Giebels},
  {Glicenstein}, {Gl{\"u}ck}, {Goret}, {G{\"o}ring}, {H{\"a}ffner}, {Hague},
  {Hampf}, {Hauser}, {Heinz}, {Heinzelmann}, {Henri}, {Hermann}, {Hinton},
  {Hoffmann}, {Hofmann}, {Hofverberg}, {Holler}, {Horns}, {Jacholkowska}, {de
  Jager}, {Jahn}, {Jamrozy}, {Jung}, {Kastendieck}, {Katarzy{\'n}ski}, {Katz},
  {Kaufmann}, {Keogh}, {Khangulyan}, {Kh{\'e}lifi}, {Klochkov}, {Klu{\.z}niak},
  {Kneiske}, {Komin}, {Kosack}, {Kossakowski}, {Laffon}, {Lamanna}, {Lennarz},
  {Lohse}, {Lopatin}, {Lu}, {Marandon}, {Marcowith}, {Masbou}, {Maurin},
  {Maxted}, {Mayer}, {McComb}, {Medina}, {M{\'e}hault}, {Moderski}, {Moulin},
  {Naumann}, {Naumann-Godo}, {de Naurois}, {Nedbal}, {Nekrassov}, {Nguyen},
  {Nicholas}, {Niemiec}, {Nolan}, {Ohm}, {de O{\~n}a Wilhelmi}, {Opitz},
  {Ostrowski}, {Oya}, {Panter}, {Paz Arribas}, {Pedaletti}, {Pelletier},
  {Petrucci}, {Pita}, {P{\"u}hlhofer}, {Punch}, {Quirrenbach}, {Raue},
  {Rayner}, {Reimer}, {Reimer}, {Renaud}, {de Los Reyes}, {Rieger}, {Ripken},
  {Rob}, {Rosier-Lees}, {Rowell}, {Rudak}, {Rulten}, {Ruppel}, {Sahakian},
  {Sanchez}, {Santangelo}, {Schlickeiser}, {Sch{\"o}ck}, {Schulz}, {Schwanke},
  {Schwarzburg}, {Schwemmer}, {Sheidaei}, {Sikora}, {Skilton}, {Sol},
  {Spengler}, {Stawarz}, {Steenkamp}, {Stegmann}, {Stinzing}, {Stycz},
  {Sushch}, {Szostek}, {Tavernet}, {Terrier}, {Tluczykont}, {Valerius}, {van
  Eldik}, {Vasileiadis}, {Venter}, {Vialle}, {Viana}, {Vincent}, {V{\"o}lk},
  {Volpe}, {Vorobiov}, {Vorster}, {Wagner}, {Ward}, {White}, {Wierzcholska},
  {Zacharias}, {Zajczyk}, {Zdziarski}, {Zech}, \& {Zechlin}}]{Abramowski12}
{Abramowski}, A., {Acero}, F., {Aharonian}, F., {et~al.} 2012, \aap, 537, A114

\bibitem[{{Acero} {et~al.}(2016){Acero}, {Ackermann}, {Ajello}, {Albert},
  {Baldini}, {Ballet}, {Barbiellini}, {Bastieri}, {Bellazzini}, {Bissaldi},
  {Bloom}, {Bonino}, {Bottacini}, {Brandt}, {Bregeon}, {Bruel}, {Buehler},
  {Buson}, {Caliandro}, {Cameron}, {Caragiulo}, {Caraveo}, {Casandjian},
  {Cavazzuti}, {Cecchi}, {Charles}, {Chekhtman}, {Chiang}, {Chiaro}, {Ciprini},
  {Claus}, {Cohen-Tanugi}, {Conrad}, {Cuoco}, {Cutini}, {D'Ammando}, {de
  Angelis}, {de Palma}, {Desiante}, {Digel}, {Di Venere}, {Drell}, {Favuzzi},
  {Fegan}, {Ferrara}, {Focke}, {Franckowiak}, {Funk}, {Fusco}, {Gargano},
  {Gasparrini}, {Giglietto}, {Giordano}, {Giroletti}, {Glanzman}, {Godfrey},
  {Grenier}, {Guiriec}, {Hadasch}, {Harding}, {Hayashi}, {Hays}, {Hewitt},
  {Hill}, {Horan}, {Hou}, {Jogler}, {J{\'o}hannesson}, {Kamae}, {Kuss},
  {Landriu}, {Larsson}, {Latronico}, {Li}, {Li}, {Longo}, {Loparco},
  {Lovellette}, {Lubrano}, {Maldera}, {Malyshev}, {Manfreda}, {Martin},
  {Mayer}, {Mazziotta}, {McEnery}, {Michelson}, {Mirabal}, {Mizuno}, {Monzani},
  {Morselli}, {Nuss}, {Ohsugi}, {Omodei}, {Orienti}, {Orlando}, {Ormes},
  {Paneque}, {Pesce-Rollins}, {Piron}, {Pivato}, {Rain{\`o}}, {Rando},
  {Razzano}, {Razzaque}, {Reimer}, {Reimer}, {Remy}, {Renault},
  {S{\'a}nchez-Conde}, {Schaal}, {Schulz}, {Sgr{\`o}}, {Siskind}, {Spada},
  {Spandre}, {Spinelli}, {Strong}, {Suson}, {Tajima}, {Takahashi}, {Thayer},
  {Thompson}, {Tibaldo}, {Tinivella}, {Torres}, {Tosti}, {Troja}, {Vianello},
  {Werner}, {Wood}, {Wood}, {Zaharijas}, \& {Zimmer}}]{fermi_diffuse}
{Acero}, F., {Ackermann}, M., {Ajello}, M., {et~al.} 2016, \apjs, 223, 26

\bibitem[{{Acero} {et~al.}(2013){Acero}, {Ackermann}, {Ajello}, {Allafort},
  {Baldini}, {Ballet}, {Barbiellini}, {Bastieri}, {Bechtol}, {Bellazzini},
  {Bland ford}, {Bloom}, {Bonamente}, {Bottacini}, {Brandt}, {Bregeon},
  {Brigida}, {Bruel}, {Buehler}, {Buson}, {Caliandro}, {Cameron}, {Caraveo},
  {Cecchi}, {Charles}, {Chaves}, {Chekhtman}, {Chiang}, {Chiaro}, {Ciprini},
  {Claus}, {Cohen-Tanugi}, {Conrad}, {Cutini}, {Dalton}, {D'Ammando}, {de
  Palma}, {Dermer}, {Di Venere}, {Silva}, {Drell}, {Drlica-Wagner}, {Falletti},
  {Favuzzi}, {Fegan}, {Ferrara}, {Focke}, {Franckowiak}, {Fukazawa}, {Funk},
  {Fusco}, {Gargano}, {Gasparrini}, {Giglietto}, {Giordano}, {Giroletti},
  {Glanzman}, {Godfrey}, {Gr{\'e}goire}, {Grenier}, {Grondin}, {Grove},
  {Guiriec}, {Hadasch}, {Hanabata}, {Harding}, {Hayashida}, {Hayashi}, {Hays},
  {Hewitt}, {Hill}, {Horan}, {Hou}, {Hughes}, {Inoue}, {Jackson}, {Jogler},
  {J{\'o}hannesson}, {Johnson}, {Kamae}, {Kawano}, {Kerr}, {Kn{\"o}dlseder},
  {Kuss}, {Lande}, {Larsson}, {Latronico}, {Lemoine-Goumard}, {Longo},
  {Loparco}, {Lovellette}, {Lubrano}, {Marelli}, {Massaro}, {Mayer},
  {Mazziotta}, {McEnery}, {Mehault}, {Michelson}, {Mitthumsiri}, {Mizuno},
  {Monte}, {Monzani}, {Morselli}, {Moskalenko}, {Murgia}, {Nakamori}, {Nemmen},
  {Nuss}, {Ohsugi}, {Okumura}, {Orienti}, {Orlando}, {Ormes}, {Paneque},
  {Panetta}, {Perkins}, {Pesce-Rollins}, {Piron}, {Pivato}, {Porter},
  {Rain{\`o}}, {Rando}, {Razzano}, {Reimer}, {Reimer}, {Reposeur}, {Ritz},
  {Roth}, {Rousseau}, {Saz Parkinson}, {Schulz}, {Sgr{\`o}}, {Siskind},
  {Smith}, {Spandre}, {Spinelli}, {Suson}, {Takahashi}, {Takeuchi}, {Thayer},
  {Thayer}, {Thompson}, {Tibaldo}, {Tibolla}, {Tinivella}, {Torres}, {Tosti},
  {Troja}, {Uchiyama}, {Vandenbroucke}, {Vasileiou}, {Vianello}, {Vitale},
  {Werner}, {Winer}, {Wood}, \& {Yang}}]{fermi_pwn}
{Acero}, F., {Ackermann}, M., {Ajello}, M., {et~al.} 2013, \apj, 773, 77

\bibitem[{{Ackermann} {et~al.}(2011){Ackermann}, {Ajello}, {Allafort},
  {Baldini}, {Ballet}, {Barbiellini}, {Bastieri}, {Belfiore}, {Bellazzini},
  {Berenji}, {Blandford}, {Bloom}, {Bonamente}, {Borgland}, {Bottacini},
  {Brigida}, {Bruel}, {Buehler}, {Buson}, {Caliandro}, {Cameron}, {Caraveo},
  {Casandjian}, {Cecchi}, {Chekhtman}, {Cheung}, {Chiang}, {Ciprini}, {Claus},
  {Cohen-Tanugi}, {de Angelis}, {de Palma}, {Dermer}, {do Couto e Silva},
  {Drell}, {Dumora}, {Favuzzi}, {Fegan}, {Focke}, {Fortin}, {Fukazawa},
  {Fusco}, {Gargano}, {Germani}, {Giglietto}, {Giordano}, {Giroletti},
  {Glanzman}, {Godfrey}, {Grenier}, {Guillemot}, {Guiriec}, {Hadasch},
  {Hanabata}, {Harding}, {Hayashida}, {Hayashi}, {Hays}, {J{\'o}hannesson},
  {Johnson}, {Kamae}, {Katagiri}, {Kataoka}, {Kerr}, {Kn{\"o}dlseder}, {Kuss},
  {Lande}, {Latronico}, {Lee}, {Longo}, {Loparco}, {Lott}, {Lovellette},
  {Lubrano}, {Martin}, {Mazziotta}, {McEnery}, {Mehault}, {Michelson},
  {Mitthumsiri}, {Mizuno}, {Monte}, {Monzani}, {Morselli}, {Moskalenko},
  {Murgia}, {Naumann-Godo}, {Nolan}, {Norris}, {Nuss}, {Ohsugi}, {Okumura},
  {Orlando}, {Ormes}, {Ozaki}, {Paneque}, {Parent}, {Pesce-Rollins},
  {Pierbattista}, {Piron}, {Pohl}, {Prokhorov}, {Rain{\`o}}, {Rando},
  {Razzano}, {Reposeur}, {Ritz}, {Parkinson}, {Sgr{\`o}}, {Siskind}, {Smith},
  {Spinelli}, {Strong}, {Takahashi}, {Tanaka}, {Thayer}, {Thayer}, {Thompson},
  {Tibaldo}, {Torres}, {Tosti}, {Tramacere}, {Troja}, {Uchiyama},
  {Vandenbroucke}, {Vasileiou}, {Vianello}, {Vitale}, {Waite}, {Wang}, {Winer},
  {Wood}, {Yang}, {Zimmer}, \& {Bontemps}}]{Ackermann11}
{Ackermann}, M., {Ajello}, M., {Allafort}, A., {et~al.} 2011, Science, 334,
  1103

\bibitem[{Ackermann {et~al.}(2012)Ackermann, Ajello, Allafort, Baldini, Ballet,
  Bastieri, Bechtol, Bellazzini, Berenji, Bloom, Bonamente, Borgland, Bouvier,
  Bregeon, Brigida, Bruel, Buehler, Buson, Caliandro, Cameron, Caraveo,
  Casandjian, Cecchi, Charles, Chekhtman, Cheung, Chiang, Cillis, Ciprini,
  Claus, Cohen-Tanugi, Conrad, Cutini, de~Palma, Dermer, Digel, do~Couto~e
  Silva, Drell, Drlica-Wagner, Favuzzi, Fegan, Fortin, Fukazawa, Funk, Fusco,
  Gargano, Gasparrini, Germani, Giglietto, Giordano, Glanzman, Godfrey,
  Grenier, Guiriec, Gustafsson, Hadasch, Hayashida, Hays, Hughes,
  J{\'{o}}hannesson, Johnson, Kamae, Katagiri, Kataoka, Knödlseder, Kuss,
  Lande, Longo, Loparco, Lott, Lovellette, Lubrano, Madejski, Martin,
  Mazziotta, McEnery, Michelson, Mizuno, Monte, Monzani, Morselli, Moskalenko,
  Murgia, Nishino, Norris, Nuss, Ohno, Ohsugi, Okumura, Omodei, Orlando, Ozaki,
  Parent, Persic, Pesce-Rollins, Petrosian, Pierbattista, Piron, Pivato,
  Porter, Rain{\`{o}}, Rando, Razzano, Reimer, Reimer, Ritz, Roth, Sbarra,
  Sgr{\`{o}}, Siskind, Spandre, Spinelli, Stawarz, Strong, Takahashi, Tanaka,
  Thayer, Tibaldo, Tinivella, Torres, Tosti, Troja, Uchiyama, Vandenbroucke,
  Vianello, Vitale, Waite, Wood, \& Yang}]{fermi_galaxy}
Ackermann, M., Ajello, M., Allafort, A., {et~al.} 2012, The Astrophysical
  Journal, 755, 164

\bibitem[{{Aharonian} {et~al.}(2018){Aharonian}, {Peron}, {Yang}, {Casanova},
  \& {Zanin}}]{aharonian20}
{Aharonian}, F., {Peron}, G., {Yang}, R., {Casanova}, S., \& {Zanin}, R. 2018,
  arXiv e-prints, arXiv:1811.12118

\bibitem[{{Aharonian} {et~al.}(2019){Aharonian}, {Yang}, \& {de O{\~n}a
  Wilhelmi}}]{aharonian19}
{Aharonian}, F., {Yang}, R., \& {de O{\~n}a Wilhelmi}, E. 2019, Nature
  Astronomy, 3, 561

\bibitem[{{Beccari} {et~al.}(2010){Beccari}, {Spezzi}, {De Marchi}, {Paresce},
  {Young}, {Andersen}, {Panagia}, {Balick}, {Bond}, {Calzetti}, {Carollo},
  {Disney}, {Dopita}, {Frogel}, {Hall}, {Holtzman}, {Kimble}, {McCarthy},
  {O'Connell}, {Saha}, {Silk}, {Trauger}, {Walker}, {Whitmore}, \&
  {Windhorst}}]{beccari10}
{Beccari}, G., {Spezzi}, L., {De Marchi}, G., {et~al.} 2010, \apj, 720, 1108

\bibitem[{{Beuther} {et~al.}(2016){Beuther}, {Bihr}, {Rugel}, {Johnston},
  {Wang}, {Walter}, {Brunthaler}, {Walsh}, {Ott}, {Stil}, {Henning},
  {Schierhuber}, {Kainulainen}, {Heyer}, {Goldsmith}, {Anderson}, {Longmore},
  {Klessen}, {Glover}, {Urquhart}, {Plume}, {Ragan}, {Schneider},
  {McClure-Griffiths}, {Menten}, {Smith}, {Roy}, {Shanahan}, {Nguyen-Luong}, \&
  {Bigiel}}]{beuther2016}
{Beuther}, H., {Bihr}, S., {Rugel}, M., {et~al.} 2016, \aap, 595, A32

\bibitem[{{Bihr} {et~al.}(2015{\natexlab{a}}){Bihr}, {Beuther}, {Ott},
  {Johnston}, {Brunthaler}, {Anderson}, {Bigiel}, {Carlhoff}, {Churchwell},
  {Glover}, {Goldsmith}, {Heitsch}, {Henning}, {Heyer}, {Hill}, {Hughes},
  {Klessen}, {Linz}, {Longmore}, {McClure-Griffiths}, {Menten}, {Motte},
  {Nguyen-Luong}, {Plume}, {Ragan}, {Roy}, {Schilke}, {Schneider}, {Smith},
  {Stil}, {Urquhart}, {Walsh}, \& {Walter}}]{bihr15}
{Bihr}, S., {Beuther}, H., {Ott}, J., {et~al.} 2015{\natexlab{a}}, \aap, 580,
  A112

\bibitem[{{Bihr} {et~al.}(2015{\natexlab{b}}){Bihr}, {Beuther}, {Ott},
  {Johnston}, {Brunthaler}, {Anderson}, {Bigiel}, {Carlhoff}, {Churchwell},
  {Glover}, {Goldsmith}, {Heitsch}, {Henning}, {Heyer}, {Hill}, {Hughes},
  {Klessen}, {Linz}, {Longmore}, {McClure-Griffiths}, {Menten}, {Motte},
  {Nguyen-Luong}, {Plume}, {Ragan}, {Roy}, {Schilke}, {Schneider}, {Smith},
  {Stil}, {Urquhart}, {Walsh}, \& {Walter}}]{Bihr2015}
{Bihr}, S., {Beuther}, H., {Ott}, J., {et~al.} 2015{\natexlab{b}}, \aap, 580,
  A112

\bibitem[{{Blasi}(2013)}]{Blasi13}
{Blasi}, P. 2013, \aapr, 21, 70

\bibitem[{{Blum} {et~al.}(1999){Blum}, {Damineli}, \& {Conti}}]{blum99}
{Blum}, R.~D., {Damineli}, A., \& {Conti}, P.~S. 1999, \aj, 117, 1392

\bibitem[{{Casandjian}(2015)}]{casandjian15}
{Casandjian}, J.-M. 2015, ArXiv e-prints arXiv:1502.07210

\bibitem[{{Case} \& {Bhattacharya}(1998)}]{case98}
{Case}, G.~L. \& {Bhattacharya}, D. 1998, \apj, 504, 761

\bibitem[{{Domingo-Santamar{\'\i}a} \& {Torres}(2006)}]{santamar06}
{Domingo-Santamar{\'\i}a}, E. \& {Torres}, D.~F. 2006, \aap, 448, 613

\bibitem[{{Ergin} {et~al.}(2014){Ergin}, {Sezer}, {Saha}, {Majumdar},
  {Chatterjee}, {Bayirli}, \& {Ercan}}]{ergin14}
{Ergin}, T., {Sezer}, A., {Saha}, L., {et~al.} 2014, \apj, 790, 65

\bibitem[{{Feng} {et~al.}(2016){Feng}, {Beuther}, {Zhang}, {Henning}, {Linz},
  {Ragan}, \& {Smith}}]{Feng2016}
{Feng}, S., {Beuther}, H., {Zhang}, Q., {et~al.} 2016, \aap, 592, A21

\bibitem[{{Gabici} {et~al.}(2007){Gabici}, {Aharonian}, \& {Blasi}}]{gabici07}
{Gabici}, S., {Aharonian}, F.~A., \& {Blasi}, P. 2007, \apss, 309, 365

\bibitem[{{Giannetti} {et~al.}(2014){Giannetti}, {Wyrowski}, {Brand},
  {Csengeri}, {Fontani}, {Walmsley}, {Nguyen Luong}, {Beuther}, {Schuller},
  {G{\"u}sten}, \& {Menten}}]{Giannetti2014}
{Giannetti}, A., {Wyrowski}, F., {Brand}, J., {et~al.} 2014, \aap, 570, A65

\bibitem[{{H.~E.~S.~S. Collaboration} {et~al.}(2018){H.~E.~S.~S.
  Collaboration}, {Abdalla}, {Abramowski}, {Aharonian}, {Ait Benkhali},
  {Ang{\"u}ner}, {Arakawa}, {Arrieta}, {Aubert}, {Backes}, {Balzer}, {Barnard},
  {Becherini}, {Becker Tjus}, {Berge}, {Bernhard}, {Bernl{\"o}hr}, {Blackwell},
  {B{\"o}ttcher}, {Boisson}, {Bolmont}, {Bonnefoy}, {Bordas}, {Bregeon},
  {Brun}, {Brun}, {Bryan}, {B{\"u}chele}, {Bulik}, {Capasso}, {Carrigan},
  {Caroff}, {Carosi}, {Casanova}, {Cerruti}, {Chakraborty}, {Chaves}, {Chen},
  {Chevalier}, {Colafrancesco}, {Condon}, {Conrad}, {Davids}, {Decock}, {Deil},
  {Devin}, {deWilt}, {Dirson}, {Djannati-Ata{\"\i}}, {Domainko}, {Donath},
  {Drury}, {Dutson}, {Dyks}, {Edwards}, {Egberts}, {Eger}, {Emery},
  {Ernenwein}, {Eschbach}, {Farnier}, {Fegan}, {Fernand es}, {Fiasson},
  {Fontaine}, {F{\"o}rster}, {Funk}, {F{\"u}{\ss}ling}, {Gabici}, {Gallant},
  {Garrigoux}, {Gast}, {Gat{\'e}}, {Giavitto}, {Giebels}, {Glawion},
  {Glicenstein}, {Gottschall}, {Grondin}, {Hahn}, {Haupt}, {Hawkes},
  {Heinzelmann}, {Henri}, {Hermann}, {Hinton}, {Hofmann}, {Hoischen}, {Holch},
  {Holler}, {Horns}, {Ivascenko}, {Iwasaki}, {Jacholkowska}, {Jamrozy},
  {Jankowsky}, {Jankowsky}, {Jingo}, {Jouvin}, {Jung-Richardt}, {Kastendieck},
  {Katarzy{\'n}ski}, {Katsuragawa}, {Katz}, {Kerszberg}, {Khangulyan},
  {Kh{\'e}lifi}, {King}, {Klepser}, {Klochkov}, {Klu{\'z}niak}, {Komin},
  {Kosack}, {Krakau}, {Kraus}, {Kr{\"u}ger}, {Laffon}, {Lamanna}, {Lau},
  {Lees}, {Lefaucheur}, {Lemi{\`e}re}, {Lemoine-Goumard}, {Lenain}, {Leser},
  {Lohse}, {Lorentz}, {Liu}, {L{\'o}pez-Coto}, {Lypova}, {Marandon},
  {Malyshev}, {Marcowith}, {Mariaud}, {Marx}, {Maurin}, {Maxted}, {Mayer},
  {Meintjes}, {Meyer}, {Mitchell}, {Moderski}, {Mohamed}, {Mohrmann},
  {Mor{\r{a}}}, {Moulin}, {Murach}, {Nakashima}, {de Naurois}, {Ndiyavala},
  {Niederwanger}, {Niemiec}, {Oakes}, {O'Brien}, {Odaka}, {Ohm}, {Ostrowski},
  {Oya}, {Padovani}, {Panter}, {Parsons}, {Paz Arribas}, {Pekeur}, {Pelletier},
  {Perennes}, {Petrucci}, {Peyaud}, {Piel}, {Pita}, {Poireau}, {Poon},
  {Prokhorov}, {Prokoph}, {P{\"u}hlhofer}, {Punch}, {Quirrenbach}, {Raab},
  {Rauth}, {Reimer}, {Reimer}, {Renaud}, {de los Reyes}, {Rieger}, {Rinchiuso},
  {Romoli}, {Rowell}, {Rudak}, {Rulten}, {Safi-Harb}, {Sahakian}, {Saito},
  {Sanchez}, {Santangelo}, {Sasaki}, {Schand ri}, {Schlickeiser},
  {Sch{\"u}ssler}, {Schulz}, {Schwanke}, {Schwemmer}, {Seglar-Arroyo},
  {Settimo}, {Seyffert}, {Shafi}, {Shilon}, {Shiningayamwe}, {Simoni}, {Sol},
  {Spanier}, {Spir-Jacob}, {Stawarz}, {Steenkamp}, {Stegmann}, {Steppa},
  {Sushch}, {Takahashi}, {Tavernet}, {Tavernier}, {Taylor}, {Terrier},
  {Tibaldo}, {Tiziani}, {Tluczykont}, {Trichard}, {Tsirou}, {Tsuji}, {Tuffs},
  {Uchiyama}, {van der Walt}, {van Eldik}, {van Rensburg}, {van Soelen},
  {Vasileiadis}, {Veh}, {Venter}, {Viana}, {Vincent}, {Vink}, {Voisin},
  {V{\"o}lk}, {Vuillaume}, {Wadiasingh}, {Wagner}, {Wagner}, {Wagner}, {White},
  {Wierzcholska}, {Willmann}, {W{\"o}rnlein}, {Wouters}, {Yang}, {Zaborov},
  {Zacharias}, {Zanin}, {Zdziarski}, {Zech}, {Zefi}, {Ziegler}, {Zorn}, \&
  {{\.Z}ywucka}}]{hgps}
{H.~E.~S.~S. Collaboration}, {Abdalla}, H., {Abramowski}, A., {et~al.} 2018,
  \aap, 612, A1

\bibitem[{{Harayama} {et~al.}(2008){Harayama}, {Eisenhauer}, \&
  {Martins}}]{harayama08}
{Harayama}, Y., {Eisenhauer}, F., \& {Martins}, F. 2008, \apj, 675, 1319

\bibitem[{{H.E.S.S.~Collaboration} {et~al.}(2015){H.E.S.S.~Collaboration},
  {Abramowski}, {Aharonian}, {Ait Benkhali}, {Akhperjanian}, {Ang{\"u}ner},
  {Backes}, {Balenderan}, {Balzer}, {Barnacka}, \& et~al.}]{Abramowski15}
{H.E.S.S.~Collaboration}, {Abramowski}, A., {Aharonian}, F., {et~al.} 2015,
  Science, 347, 406

\bibitem[{{Jackson} {et~al.}(2006){Jackson}, {Rathborne}, {Shah}, {Simon},
  {Bania}, {Clemens}, {Chambers}, {Johnson}, {Dormody}, {Lavoie}, \&
  {Heyer}}]{Jackson2006}
{Jackson}, J.~M., {Rathborne}, J.~M., {Shah}, R.~Y., {et~al.} 2006, \apjs, 163,
  145

\bibitem[{{Lande} {et~al.}(2012){Lande}, {Ackermann}, {Allafort}, {Ballet},
  {Bechtol}, {Burnett}, {Cohen-Tanugi}, {Drlica-Wagner}, {Funk}, {Giordano},
  {Grondin}, {Kerr}, \& {Lemoine-Goumard}}]{Lande12}
{Lande}, J., {Ackermann}, M., {Allafort}, A., {et~al.} 2012, \apj, 756, 5

\bibitem[{{Manchester} {et~al.}(2005){Manchester}, {Hobbs}, {Teoh}, \&
  {Hobbs}}]{atnf}
{Manchester}, R.~N., {Hobbs}, G.~B., {Teoh}, A., \& {Hobbs}, M. 2005, \aj, 129,
  1993

\bibitem[{{Motte} {et~al.}(2003){Motte}, {Schilke}, \& {Lis}}]{motte03}
{Motte}, F., {Schilke}, P., \& {Lis}, D.~C. 2003, \apj, 582, 277

\bibitem[{{Nguyen Luong} {et~al.}(2011){Nguyen Luong}, {Motte}, {Schuller},
  {Schneider}, {Bontemps}, {Schilke}, {Menten}, {Heitsch}, {Wyrowski},
  {Carlhoff}, {Bronfman}, \& {Henning}}]{luong11}
{Nguyen Luong}, Q., {Motte}, F., {Schuller}, F., {et~al.} 2011, \aap, 529, A41

\bibitem[{{Nguyen-Luong} {et~al.}(2016){Nguyen-Luong}, {Nguyen}, {Motte},
  {Schneider}, {Fujii}, {Louvet}, {Hill}, {Sanhueza}, {Chibueze}, \&
  {Didelon}}]{luong16}
{Nguyen-Luong}, Q., {Nguyen}, H. V.~V., {Motte}, F., {et~al.} 2016, \apj, 833,
  23

\bibitem[{{Papadopoulos}(2010)}]{papadopoulos10}
{Papadopoulos}, P.~P. 2010, \apj, 720, 226

\bibitem[{{Roman-Duval} {et~al.}(2016){Roman-Duval}, {Heyer}, {Brunt}, {Clark},
  {Klessen}, \& {Shetty}}]{Roman-Duval2016}
{Roman-Duval}, J., {Heyer}, M., {Brunt}, C.~M., {et~al.} 2016, \apj, 818, 144

\bibitem[{{Roman-Lopes} {et~al.}(2018){Roman-Lopes},
  {Rom{\'a}n-Z{\'u}{\~n}iga}, {Tapia}, {Chojnowski}, {G{\'o}mez Maqueo Chew},
  {Garc{\'\i}a-Hern{\'a}ndez}, {Borissova}, {Minniti}, {Covey},
  {Longa-Pe{\~n}a}, {Fernandez-Trincado}, {Zamora}, \& {Nitschelm}}]{apogee1}
{Roman-Lopes}, A., {Rom{\'a}n-Z{\'u}{\~n}iga}, C., {Tapia}, M., {et~al.} 2018,
  \apj, 855, 68

\bibitem[{{Roman-Lopes} {et~al.}(2019){Roman-Lopes},
  {Rom{\'a}n-Z{\'u}{\~n}iga}, {Tapia}, {Hern{\'a}ndez},
  {Ram{\'\i}rez-Preciado}, {Stringfellow}, {Ybarra}, {Kim}, {Minniti}, {Covey},
  {Kounkel}, {Su{\'a}rez}, {Borissova}, {Garc{\'\i}a-Hern{\'a}ndez}, {Zamora},
  \& {Trujillo}}]{apogee2}
{Roman-Lopes}, A., {Rom{\'a}n-Z{\'u}{\~n}iga}, C.~G., {Tapia}, M., {et~al.}
  2019, \apj, 873, 66

\bibitem[{{Stil} {et~al.}(2006){Stil}, {Taylor}, {Dickey}, {Kavars}, {Martin},
  {Rothwell}, {Boothroyd}, {Lockman}, \& {McClure-Griffiths}}]{stil2006}
{Stil}, J.~M., {Taylor}, A.~R., {Dickey}, J.~M., {et~al.} 2006, \aj, 132, 1158

\bibitem[{{The Fermi-LAT collaboration}(2019)}]{Fermi19}
{The Fermi-LAT collaboration}. 2019, arXiv e-prints
  arXiv:1902.10045

\bibitem[{{Umemoto} {et~al.}(2017){Umemoto}, {Minamidani}, {Kuno}, {Fujita},
  {Matsuo}, {Nishimura}, {Torii}, {Tosaki}, {Kohno}, {Kuriki}, {Tsuda},
  {Hirota}, {Ohashi}, {Yamagishi}, {Handa}, {Nakanishi}, {Omodaka}, {Koide},
  {Matsumoto}, {Onishi}, {Tokuda}, {Seta}, {Kobayashi}, {Tachihara}, {Sano},
  {Hattori}, {Onodera}, {Oasa}, {Kamegai}, {Tsuboi}, {Sofue}, {Higuchi},
  {Chibueze}, {Mizuno}, {Honma}, {Muller}, {Inoue}, {Morokuma-Matsui},
  {Shinnaga}, {Ozawa}, {Takahashi}, {Yoshiike}, {Costes}, \&
  {Kuwahara}}]{Umemoto2017}
{Umemoto}, T., {Minamidani}, T., {Kuno}, N., {et~al.} 2017, Publications of the
  Astronomical Society of Japan, 69, 78

\bibitem[{{Wang} {et~al.}(2019){Wang}, {Beuther}, {Rugel}, {Soler}, {Stil},
  {Ott}, {Bihr}, {McClure-Griffiths}, {Anderson}, {Klessen}, {Goldsmith},
  {Roy}, {Glover}, {Urquhart}, {Heyer}, {Linz}, {Smith}, {Bigiel}, {Dempsey},
  \& {Henning}}]{Wang2019}
{Wang}, Y., {Beuther}, H., {Rugel}, M.~R., {et~al.} 2019, arXiv e-prints,
  arXiv:1912.08223

\bibitem[{{Wang} {et~al.}(2018){Wang}, {Bihr}, {Rugel}, {Beuther}, {Johnston},
  {Ott}, {Soler}, {Brunthaler}, {Anderson}, {Urquhart}, {Klessen}, {Linz},
  {McClure-Griffiths}, {Glover}, {Menten}, {Bigiel}, {Hoare}, \&
  {Longmore}}]{Wang2018}
{Wang}, Y., {Bihr}, S., {Rugel}, M., {et~al.} 2018, \aap, 619, A124

\bibitem[{{Wilson} {et~al.}(2013){Wilson}, {Rohlfs}, \&
  {H{\"u}ttemeister}}]{wilson2013}
{Wilson}, T.~L., {Rohlfs}, K., \& {H{\"u}ttemeister}, S. 2013, {Tools of Radio
  Astronomy}

\bibitem[{{Wright} {et~al.}(2010){Wright}, {Drake}, {Drew}, \&
  {Vink}}]{wright10}
{Wright}, N.~J., {Drake}, J.~J., {Drew}, J.~E., \& {Vink}, J.~S. 2010, \apj,
  713, 871

\bibitem[{{Yang} {et~al.}(2016){Yang}, {Aharonian}, \& {Evoli}}]{yang16}
{Yang}, R., {Aharonian}, F., \& {Evoli}, C. 2016, \prd, 93, 123007

\bibitem[{{Yang} \& {Aharonian}(2017)}]{Yang17}
{Yang}, R.-z. \& {Aharonian}, F. 2017, \aap, 600, A107

\bibitem[{{Yang} {et~al.}(2018){Yang}, {de O{\~n}a Wilhelmi}, \&
  {Aharonian}}]{Yang18}
{Yang}, R.-z., {de O{\~n}a Wilhelmi}, E., \& {Aharonian}, F. 2018, \aap, 611,
  A77

\bibitem[{{Zhang} {et~al.}(2014){Zhang}, {Moscadelli}, {Sato}, {Reid},
  {Menten}, {Zheng}, {Brunthaler}, {Dame}, {Xu}, \& {Immer}}]{Zhang2014}
{Zhang}, B., {Moscadelli}, L., {Sato}, M., {et~al.} 2014, \apj, 781, 89

\end{thebibliography}

\end{document}